\documentclass[
 reprint,
 amsmath,amssymb,
 aps,
 prb,
]{revtex4-2}

\usepackage{graphicx}
\usepackage{dcolumn}
\usepackage{bm}
\usepackage{braket}
\usepackage{xcolor}
\usepackage{booktabs}
\usepackage[version=3]{mhchem}
\usepackage{hyperref}
\hypersetup{
  colorlinks=true,
  linkcolor=blue,   
  citecolor=blue,  
  urlcolor=black     
}

\begin{document}

\preprint{APS/123-QED}

\title{Non-Adiabatic Spin-Induced Currents in Magnetic Weyl Semimetals}

\author{
Kota Watanabe$^{1}$, 
Akihiro Ozawa$^{2,3}$, 
Hiroaki Ishizuka$^{4}$, 
Hiroki Isobe$^{1}$, 
Kentaro Nomura$^{1,5}$ \\
\vspace{0.5em} 
\itshape\small $^{1}$Department of Physics, Kyushu University, Fukuoka 819-0395, Japan \\
\itshape\small $^{2}$Institut f\"ur Theoretische Physik, Universit\"at Leipzig, 04103 Leipzig, Germany \\
\itshape\small $^{3}$Institute for Solid State Physics, The University of Tokyo, Kashiwa 277-8581, Japan \\
\itshape\small $^{4}$Department of Physics, Institute of Science Tokyo, Meguro, Tokyo 152-8551, Japan \\
\itshape\small $^{5}$Quantum and Spacetime Research Institute, Kyushu University, Fukuoka 819-0395, Japan
}

\begin{abstract}
We theoretically study dc electric currents induced by rapid spin precession in noncentrosymmetric magnetic Weyl semimetals using the Floquet--Keldysh Green's-function formalism.
The spin dynamics act as a time-dependent chiral gauge field and the geometric structure of electric currents is expressed in terms of the Berry curvature, the quantum metric, and the shift vector.
For an isotropic Weyl cone, only an injection current is generated. This current scales quadratically with the driving frequency and linearly with the relaxation time associated with coupling to a fermionic bath.
We further show that a finite Weyl-cone tilt modifies the available interband-transition phase space, smooths the frequency dependence of the current, and allows for a shift current and another type of injection current, both of which flow in directions determined by the tilt and the direction of the precession.
A finite net current can remain after summing contributions from Weyl nodes of opposite chiralities when their transition phase spaces are inequivalent.
\end{abstract}

\maketitle

\section{Introduction}

Magnets hosting strong exchange interactions can exhibit nontrivial cross-correlated responses, such as the generation of an electric current driven by magnetization dynamics \cite{Volovik1987, Nagaosa-Tokura2012, Barnes2007}. The emergence of such cross-correlated phenomena requires a low-symmetry environment, specifically the simultaneous breaking of spatial inversion ($\cal P$) and time-reversal ($\cal T$) symmetries \cite{ Essin2010, hi2023}. The investigation of these phenomena is driven by the aim of spintronics to develop faster, more efficient, and highly precise devices \cite{Zutic2004,Puebla2020}. Achieving this goal requires a comprehensive understanding of rapid magnetization dynamics and their non-equilibrium interplay with itinerant electrons \cite{Tserkovnyak2002, VarelaManjarres2024}.

Therefore, we focus on magnetic Weyl semimetals (MWSMs) as an ideal platform for this pursuit, as they satisfy the necessary symmetry requirements while featuring strong interactions between topological conduction electrons and localized spins \cite{Burkov-Balents2011,Wan2011-ck, amv2018, Araki2020, Nagaosa-Morimoto-Tokura2020,Nikolic2021, ozawa2026magnetic}. Noncentrosymmetric MWSMs considered here lacking inversion symmetry exhibit nonlinear responses driven by magnetization, promising a structure as rich as, if not richer than, conventional optical responses \cite{Nagaosa-Morimoto-Tokura2020, hi2023, meg}. Despite this potential, existing studies on spin-induced electric currents have largely been confined to the adiabatic regime, where spin fluctuations are sufficiently slow \cite{Ishizuka2016,hi2023}. The cross-correlated electric current response in the non-adiabatic regime, where fast magnetization dynamics drive the system out of equilibrium, remains a crucial yet unexplored frontier.

In this paper, we investigate a MWSM with broken $\mathcal{P}$ symmetry driven out of equilibrium by spin dynamics to explore non-adiabatic geometric transport. Specifically, we derive a formula for the spin-induced nonlinear electric current using the Floquet--Keldysh Green's function method \cite{kohler2005driven, kamenev,oka2009photovoltaic, at2014, mn2016, Matsyshyn2021}.
This paper is organized as follows.
We first introduce a simple continuum model of a MWSM under time-periodic spin precession. 
We then derive analytical expressions for the spin-induced current using the Keldysh Green's function formalism and Floquet theory (Sec. II). 
In Sec. III, we show that the resulting current is decomposed into shift-current and injection-current contributions, which are characterized by band-topological or geometrical quantities, such as the Berry curvature, the quantum metric, and the shift vector. 
We find that tilting of the Weyl cone induces a finite current with a characteristic dependence on the driving frequency.
Then in Sec. IV, we evaluate the strength of the induced current based on the effective model of a certain MWSM, compensated ferrimagnetic Heusler system \ce{Ti2MnAl}.

\section{Method}

\subsection{Model}

\begin{figure}[tbp]
    \centering
    \includegraphics[width=1\linewidth]{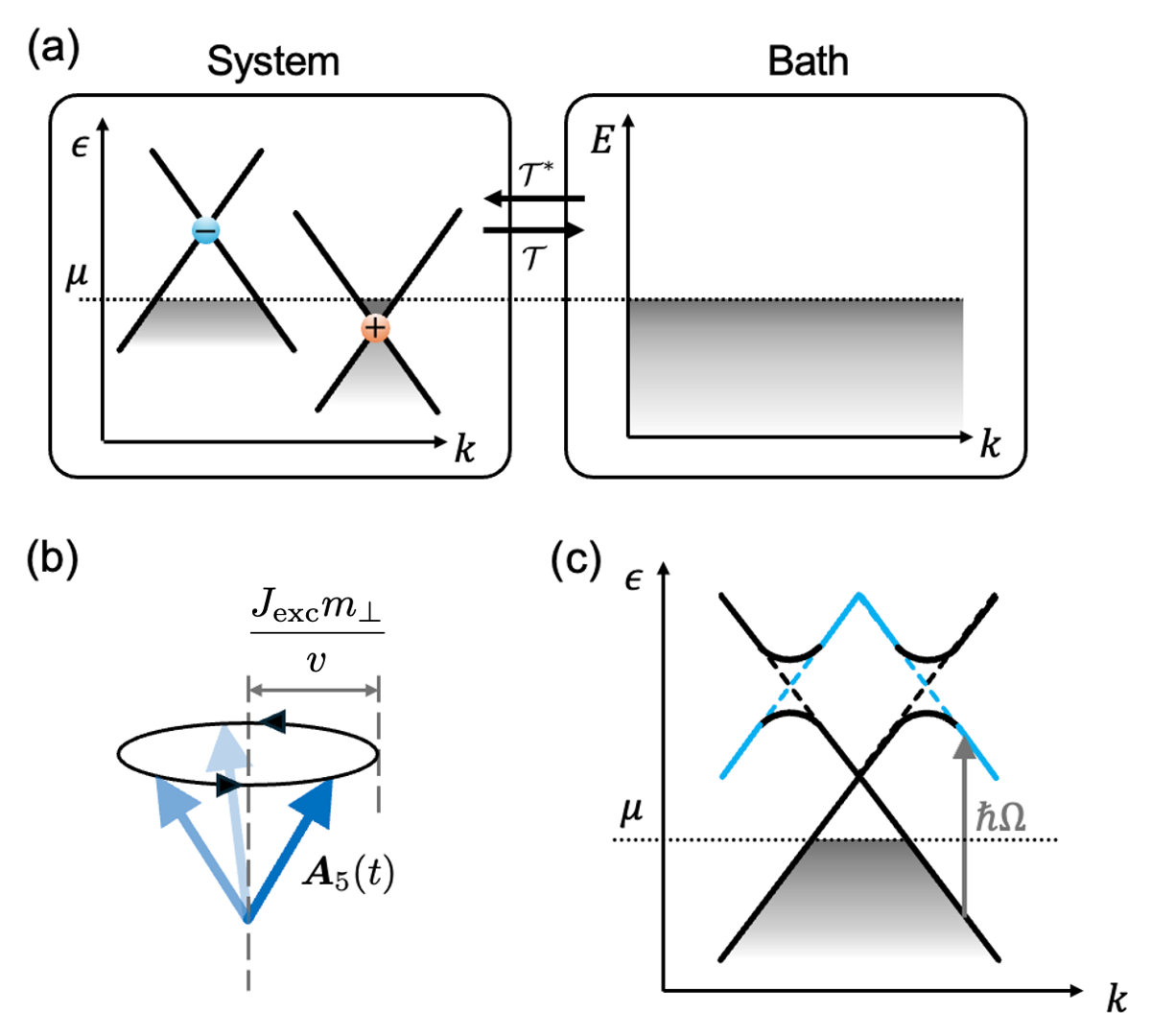}
    \caption{Model of the driven MWSM. (a) Band structure of a MWSM lacking inversion ($\mathcal{P}$) symmetry, indicating the distinct chemical potentials of opposite chirality. The tunneling coefficient between the system and the bath is denoted by $\mathcal{T}$. $\mathcal{T}^*$ for the process from the bath to the system. (b) The effective chiral gauge field $ A_5$ induced as a perturbation.(c) The relevant Floquet band sector, highlighting the resonance between the one-photon-dressed valence band and the bare conduction band.}
    \label{fig:fig1}
\end{figure}

\begin{figure}[tbp]
    \centering
    \includegraphics[width=1\linewidth]{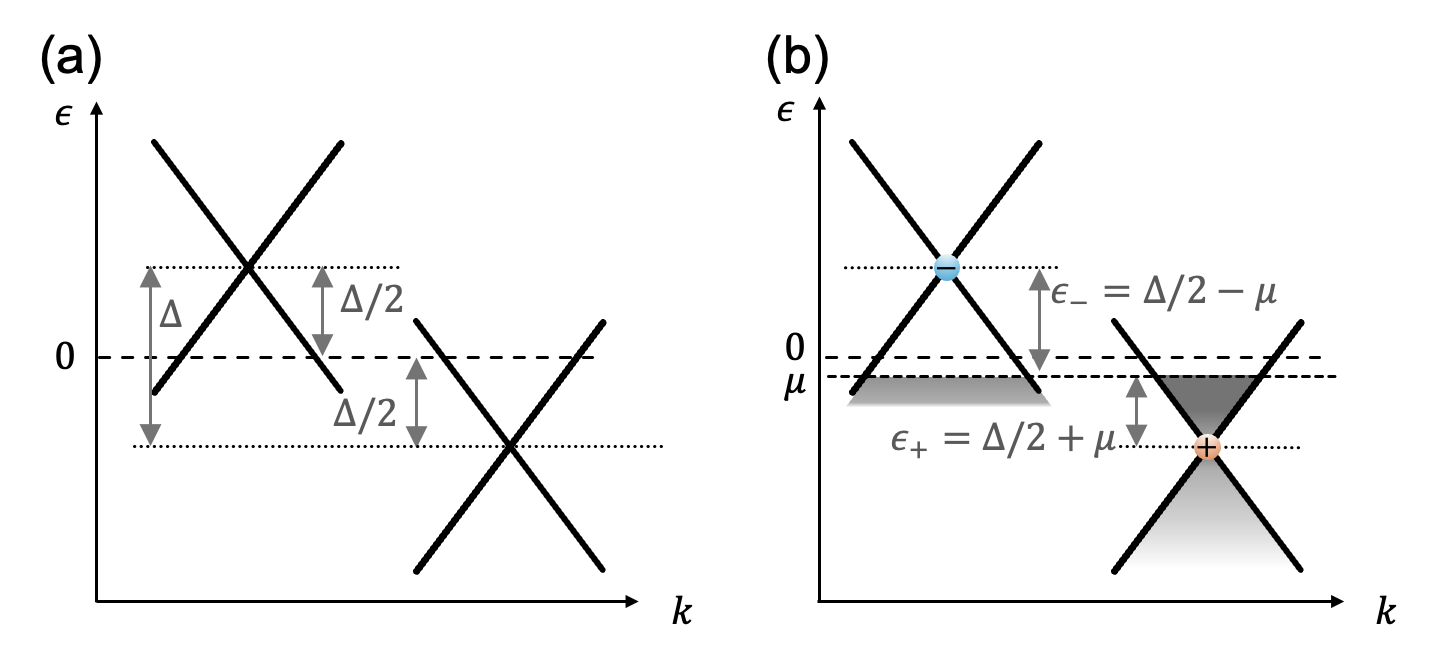}
    \caption{Schematic illustration of the energy offset between Weyl nodes. (a) The energy splitting between the two Weyl nodes is denoted by $\Delta$, with the nodes symmetrically displaced by $\pm\Delta/2$ relative to the reference energy $\epsilon = 0$. (b) Assuming the $s = -1$ Weyl node is located at a higher energy than the $s = +1$ Weyl node, the energy differences relative to the chemical potential $\mu$ are given by $\epsilon_- = \Delta/2 - \mu$ for the $s =- 1$ node and $\epsilon_+ = -(\Delta/2 + \mu)$ for the $s = +1$ Weyl node. The gray areas indicate the occupied electronic states.}
    \label{fig:weyl_nodes_position}
\end{figure}

We consider the following model of noncentrosymmetric magnetic Weyl semimetals (MWSMs) coupled to a ferromagnetic moment, whose effective Hamiltonian is given by \cite{ym2017, amv2018}
\begin{widetext}
\begin{equation}
    \mathcal{H}_{\mathrm{MWSM}} (t) =
    \begin{pmatrix}
        -i {v}\bm{\sigma}\cdot \bm{\nabla} - J_{\mathrm{exc}} \bm{m}(t)\cdot\bm{\sigma} - (\mu + \Delta/2 ) \sigma_0 & 0 \\
        0 & i {v}\bm{\sigma}\cdot \bm{\nabla} - J_{\mathrm{exc}} \bm{m}(t)\cdot\bm{\sigma} - (\mu - \Delta/2 ) \sigma_0 
    \end{pmatrix},
    \label{eq:MWSMham_original}
\end{equation}
\end{widetext}
where $v$ is the Fermi velocity, $J_{\mathrm{exc}}$ is the exchange coupling constant, and $\Delta$ denotes the energy splitting between the two Weyl nodes (see Fig. \ref{fig:fig1}-(a), \ref{fig:weyl_nodes_position}). 
To be specific, we consider a magnetic precession about the $z$ axis, in which case $\bm m(t)$ reads
\begin{align}
    \bm{m}(t) &= \delta \bm{m}(t) + \bm{m}_0, \\
    \delta \bm{m}(t) &= m_\perp (\cos \Omega t, \sin \Omega t, 0)^T, \\
    \bm{m}_0 &= \left( 0, 0, \sqrt{1-m_\perp^2} \right)^T.
\end{align}
By absorbing the static magnetization into the momentum as
$\bm k \equiv -i  \bm{\nabla} - s {J_{\rm exc}}\bm m_0 /v$,
and introducing the chiral gauge field 
$\bm A_5(t) \equiv {J_{\rm exc}}\delta\bm m(t)/ v$ \cite{pcf2016, Liang2025},
we can compactly rewrite the single node Hamiltonian as
\begin{align}
    {\cal H}(t)
    &=
    s v (\bm k- s \bm A_5(t))
    \cdot\bm{\sigma}
    - \Big(\mu + s \frac{\Delta}{2} \Big) \sigma_0 ,
    \label{eq:MWSMham}
\end{align}
where $s=\pm1$ labels the chirality of the Weyl node (see \ref{fig:fig1}-(b)).

To treat the chirality and band indices in a unified and compact manner, we parameterize the momentum vector as $\bm{k} = k (\sin\theta\cos\phi, \sin\theta\sin\phi, \cos\theta)^T$ with $k \equiv |\bm{k}|$, and define the standard eigenstates of the chirality operator $\bm{k} \cdot \bm{\sigma}$ as follows:
\begin{align}
    (\bm{k} \cdot \bm{\sigma}) \ket{\chi_\pm} &= \pm k \ket{\chi_\pm}, \notag \\
    \ket{\chi_+} = \begin{pmatrix} \cos \frac{\theta}{2} \\ e^{i\phi}\sin\frac{\theta}{2} \end{pmatrix},
    &\quad
    \ket{\chi_-} = \begin{pmatrix} \sin \frac{\theta}{2} \\ -e^{i\phi} \cos\frac{\theta}{2} \end{pmatrix}.
\end{align}
Using these basis states, the eigenvalue equation for the unperturbed Hamiltonian $\mathcal{H}_0 = s v \bm{k} \cdot \bm{\sigma} - (\mu + s {\Delta}/{2} ) \sigma_0$ can be expressed for both chiralities $s = \pm 1$ as
\begin{equation}
    \mathcal{H}_0 \ket{u_n^s} = \epsilon_n^s \ket{u_n^s}, \quad (n = 1, 2)
\end{equation}
where the band energies and their corresponding eigenspinors are explicitly given by
\begin{align}
    \epsilon_2^s 
    &= v k - \Big(\mu + s \frac{\Delta}{2} \Big), \quad \ket{u_2^s} 
    = \ket{\chi_{+s}}, \\
    \epsilon_1^s 
    &= -v k - \Big(\mu + s \frac{\Delta}{2} \Big), \quad \ket{u_1^s} 
    = \ket{\chi_{-s}},
\end{align}
with the indices $2$ and $1$ denoting the conduction and valence bands, respectively. 
Using this notation, we define the matrix elements of a generic Hermitian operator $\mathcal{O}$ as
\begin{equation}
    \mathcal{O}_{ij} \equiv \bra{u_i^s} \mathcal{O} \ket{u_j^s},
\end{equation}
where $i$ and $j$ denote the band indices.

We model dissipation by coupling the system to an equilibrium fermionic reservoir with chemical potential $\mu$ and inverse temperature $\beta>0$ (Fig. \ref{fig:fig1}(a)). Within the wide-band approximation, the reservoir hybridization is assumed to be energy independent over the relevant energy window; after absorbing the constant real part into the system Hamiltonian, the retarded self-energy reduces to a frequency-independent imaginary broadening characterized by $\Gamma$ (see Appendix~\ref{FK}). This featureless bath provides a phenomenological description of relaxation by encoding the net effect of the environment into a single parameter, and therefore $\Gamma$ should not be identified with the rate of any specific microscopic scattering mechanism \cite{Datta2000,Jauho1994,mn2016}.

\subsection{Floquet effective Hamiltonian}

Since the magnetization $\bm{m}(t)$ precesses with frequency $\Omega$, i.e. the Hamiltonian $\mathcal{H} (t)$ has the time periodicity of $T \equiv 2\pi/\Omega $, we expand the Hamiltonian into its Fourier components:
\begin{equation}
    H_n (\bm{k}) = \int_0^T \frac{dt}{T} e^{in \Omega t} \mathcal{H} (\bm{k}, t).
\end{equation}
Decomposing $\bm{A}_5 (t)=\bm{A}_{5,\Omega} e^{-i\Omega t} + \bm{A}_{5,\Omega}^* e^{i\Omega t} $, yielding
\begin{align}
    \bm{A}_{5, \Omega } &= \frac{J_{\rm exc} m_\perp}{2v} ( 1 , i, 0 )^T 
    =  \frac{A_5}{\sqrt{2}} ( 1 , i , 0 )^T  ,
\end{align}
where $ A_5 \equiv J_{\rm exc} m_\perp /(\sqrt{2} v)$, the unperturbed Hamiltonian $H_0$ and the coupling component $H_1$ are given by $H_0 = s v \bm{k} \cdot \bm{\sigma} -  (\mu + s {\Delta}/{2} ) \sigma_0 = \mathcal{H}_0$ and $H_{1} = - v \bm{A}_{5, \Omega}\cdot \bm{\sigma} , \; H_{-1} = - v\bm{A}_{5, \Omega}^* \cdot \bm{\sigma} $.
Using the Floquet formalism detailed in Appendix \ref{FK}, the effective Hamiltonian of the system is given by \cite{s1965, s1973}
\begin{align}
    (H_F)_{l, n} \equiv H_{l-n} - n\Omega\delta_{l,n}.
\end{align}

To explicitly calculate the electric current, we reduce the infinite-dimensional Floquet space to a tractable subspace \cite{mn2016}. 
Focusing on the single-photon processes, we truncate the Floquet space to the resonant subspace spanned by $n=0$ and $-1$ (see Fig. \ref{fig:fig1}(c)). 
The Floquet effective Hamiltonian in this truncated subspace is given by \cite{Fregoso2013}
\begin{align}
    H_F^{2\times 2} &=
    \begin{pmatrix}
        \epsilon_1 + \Omega & -s \frac{A_5}{\sqrt{2}} (v_{12}^x - i v_{12}^y) \\
        -s \frac{A_5}{\sqrt{2}} (v_{21}^x + i v_{21}^y) & \epsilon_2 
    \end{pmatrix} \notag \\
    &= h_0 \tau_0 + \bm{h} \cdot \bm{\tau},
    \label{eq:Floquet_eff_Ham}
\end{align}
where the velocity matrix elements $v_{ij}^a$ are defined as
$v_{ij}^a \equiv \bra{u_i^s} \partial_a \mathcal{H}_0 \ket{u_j^s}$ (see \ref{fig:fig1}-(c)).

The Floquet formalism requires the time periodicity of the Hamiltonian,
${\cal H}(t+T)={\cal H}(t)$, and therefore remains applicable to the
gapless Weyl spectrum~\cite{s1965,s1973}.
In the present system, the relevant non-adiabatic response arises from
the one-photon resonance,
$\epsilon_2-\epsilon_1\approx\hbar\Omega$.
Since $\epsilon_2-\epsilon_1=2vk$, the resonant surface is determined by
$2vk=\hbar\Omega$.
The interband transitions are further restricted by Pauli blocking,
$\hbar\Omega\gtrsim2|\mu+s\Delta/2|$.
Thus, $\Delta$ is not a band gap required for the Floquet formalism,
but it sets the characteristic energy scale of the resonant response,
$\hbar\Omega\sim\Delta$, up to the chemical-potential shift
(see Fig.~\ref{fig:fig1}-(a)).

Near the one-photon resonance, the one-photon-dressed valence band
$\ket{u_1,-1}$ and the bare conduction band $\ket{u_2,0}$ become nearly
degenerate, $\epsilon_1+\hbar\Omega\approx \epsilon_2$.
The dominant resonant subspace is therefore described by the $2\times 2 $ Floquet effective Hamiltonian (\ref{eq:Floquet_eff_Ham}).
The reduction to this two-level subspace is justified within the
rotating-wave approximation when the coupling between Floquet sectors
is sufficiently smaller than their energy separation,
$|A_{5,\Omega}\cdot\bm v_{12}|,\Gamma\ll\hbar\Omega$.
Under this condition, the off-resonant Floquet sectors give only
higher-order corrections and can be neglected
\cite{Matsyshyn2021,Fregoso2013}.
Thus, the periodicity guarantees the Floquet description, whereas the
near-resonance condition and weak intersector coupling justify the
effective $2\times2$ reduction.

Its coefficients in the Pauli matrix representation by using $\tau$, which is two-level Floquet basis, are given by
\begin{align}
    h_{0}
    &=
    \frac{(\epsilon_{1}+\Omega)+\epsilon_{2}}{2},
    \quad
    h_{z}
    =
    \frac{(\epsilon_{1}+\Omega)-\epsilon_{2}}{2},
    \\
    &h_{x}+ih_{y}
    =
    -s\frac{A_{5}}{\sqrt{2}}
    \left(v_{21}^{x}+iv_{21}^{y}\right),
    \\
    &h_{x}-ih_{y}
    =
    -s\frac{A_{5}}{\sqrt{2}}
    \left(v_{12}^{x}-iv_{12}^{y}\right).
\end{align}

Focusing on the two Floquet bands that cross, we construct an effective Hamiltonian and decompose the group velocity into identity and Pauli matrix components:
\begin{align}
    & \tilde{v}^a \equiv
    \notag \\
    &
    \begin{pmatrix}
        \partial_{a} \epsilon_1
        &
        -s\frac{A_5}{\sqrt{2}}
        ((\partial_{a} v^x)_{12} -i (\partial_{a} v^y)_{12})
        \\
        -s\frac{A_5}{\sqrt{2}}
        ((\partial_{a} v^x) _{21}+i (\partial_{a}  v^y)_{21})
        &
        \partial_{a} \epsilon_2
    \end{pmatrix}
    \notag \\
    &= \tilde{v}_0^a \tau_0
    + \tilde{\bm v}^{a}\cdot\bm{\tau},
\end{align}
with the coefficients given by
\begin{align}
    \tilde{v}_{0}^{a}
    &= \frac{v^{a}_{11}+v^{a}_{22}}{2},
    \quad
    \tilde{v}_{z}^{a}
    = \frac{v^{a}_{11}-v^{a}_{22}}{2},
    \\
    \tilde{v}_{x}^{a}+i\tilde{v}_{y}^{a}
    &=
    -s\frac{A_{5}}{\sqrt{2}}
    \left[
        (\partial_{a}v^{x})_{21}
        +i(\partial_{a}v^{y})_{21}
    \right],
    \\
    \tilde{v}_{x}^{a}-i\tilde{v}_{y}^{a}
    &=
    -s\frac{A_{5}}{\sqrt{2}}
    \left[
        (\partial_{a}v^{x})_{12}
        -i(\partial_{a}v^{y})_{12}
    \right],
\end{align}
and we use the notation
$\partial_a \equiv \partial/\partial k_a$.
Within this two-level Floquet basis $\bm{\tau}$,
the lesser component of the self-energy in frequency space takes a diagonal form because of the wide-band approximation:
\begin{equation}
    \bm{\Sigma}^<(\omega)
    =
    i\Gamma
    \begin{pmatrix}
        f(\omega - \Omega) & 0 \\
        0             & f(\omega)
    \end{pmatrix},
\end{equation}
where $\tau = 1/(2\Gamma)$ represents the relaxation time, and $f(\omega)$ denotes the Fermi distribution function of the bath. The bold indicates the matrix of the $2\times 2$ Floquet basis.

To evaluate the lesser Green's function via the Dyson equation,
\begin{equation}
    \bm{G}^<(\omega) = \bm{G}^R(\omega)\bm{\Sigma}^<(\omega)\bm{G}^A(\omega),
\end{equation}
we first require the retarded and advanced Green's functions (see Appendix \ref{FK}). Within the truncated two-level Floquet subspace, these can be explicitly determined from the reduced Floquet effective Hamiltonian $H_F^{2\times 2}$ as follows:
\begin{equation}
    \bm{G}^{R/A}(\omega) = \frac{(\omega - h_0 \pm i\Gamma/2)\sigma_0 + \bm{h} \cdot \bm{\tau}}{(\omega - h_0 \pm i\Gamma/2)^2 - h^2},
\end{equation}
where $h \equiv |\bm{h}|, \; \bm{h}\equiv(h_x,h_y,h_z)^T$.

\section{Results}

The $a$-component of the electric current density induced by the magnetization precession in the $xy$ plane is determined by the lesser Green's function as
\begin{equation}
    \bar J^{a;xy}
    \equiv
    - i 
    \int \frac{d\bm{k}}{(2\pi)^3}
    \int \frac{d\omega}{2\pi}
    \operatorname{Tr}
    \Big[
    \tilde{v}^a(\bm{k})
    \bm{G}^{<}(\bm{k},\omega)
\Big],
\label{eq:dc}
\end{equation}
where $d \bm k \equiv dk_x dk_y dk_z$.

Performing the frequency integration and extracting the time-independent component, the total DC electric current can be decomposed into two contributions.
Eq.~(\ref{eq:dc}) reduces to
\begin{align}
    \bar J^{a;xy}
    = \bar J_{\rm shift}^{a;xy}+\bar J_{\rm inj}^{a;xy},
\end{align}
where
\begin{align}
    \bar J_{\rm shift}^{a;xy}
    &\equiv
    \int \frac{d\bm{k}}{(2\pi)^3}
    \frac{\Gamma}{4}
    \frac{[\tilde{\bm{v}}^a\times\bm{h}]_z}{h^2+\Gamma^2/4}
    \delta f,
    \label{eq:j1_weak}
    \\
    \bar J_{\rm inj}^{a;xy}
    &\equiv
    \int \frac{d\bm{k}}{(2\pi)^3}
    \frac{1}{2}
    \frac{h_z(\tilde{\bm{v}}^a\cdot\bm{h})+\tilde v_z^a \Gamma^2/4}{h^2+\Gamma^2/4}
    \delta f.
    \label{eq:j3_weak}
\end{align}
Here, we have introduced the combination of Fermi distribution functions
$\delta f \equiv f(h_0+h-\Omega)+f(h_0-h-\Omega)-f(h_0+h)-f(h_0-h)$, following Ref. \cite{mn2016}, and defined
$\bm{h}\equiv(h_x,h_y,h_z)^T,\;\tilde{\bm{v}}^a\equiv(\tilde v_x^a,\tilde v_y^a,\tilde v_z^a)^T$.
We classify these two contributions according to their dependence on the relaxation rate $\Gamma$ and refer to them as the shift and injection currents, respectively.
The physical origin of this correspondence will be clarified below.

\subsection{Perturbative Limit}

We now consider the perturbative regime
$|\bm A_{5,\Omega}\cdot\bm v_{12}|\ll\Gamma$, where the off-diagonal hybridization amplitude of the Floquet effective Hamiltonian is much smaller than the relaxation rate $\Gamma$ induced by the coupling to the bath.
When the resonance width is also sufficiently small compared with the characteristic energy scale of the interband transition, the Lorentzian factor associated with spectral broadening becomes sharply peaked near resonance.
It can then be approximated by a Dirac delta function as
\begin{align}
	\frac{1}{h^2+\Gamma^2/4}
	&\approx
	\frac{4\pi}{\Gamma}
	\delta(\epsilon_{12}+\Omega).
	\label{eq:approx}
\end{align}

Here, the delta-function form should be understood as a narrow-resonance approximation.
More generally, retaining the drive-induced hybridization $\sqrt{h_x^2+h_y^2}$ gives \cite{Matsyshyn2021}
\begin{align}
	\frac{1}{h_z^2+h_x^2+h_y^2+\Gamma^2/4}
	&\approx
	\frac{\pi}{\sqrt{h_x^2+h_y^2+\Gamma^2/4}}
	\delta(h_z).
\end{align}
In the perturbative regime $\sqrt{h_x^2+h_y^2}\ll\Gamma$, this expression reduces to Eq.~\eqref{eq:approx}, with corrections of order $O[(h_x^2+h_y^2)/\Gamma^2]$.
For the Weyl model considered here, the present resonance-shell approximation is therefore justified under the hierarchy
$\sqrt{h_x^2+h_y^2}\ll\Gamma\ll\Omega$.
Note that for $\sqrt{h_x^2+h_y^2}\gtrsim\Gamma$, nonperturbative Rabi saturation becomes important and modifies both the magnitude and the relaxation-time dependence of the current \cite{Matsyshyn2021}.

Using Eq.~\eqref{eq:approx}, the resonance condition selects the interband transitions satisfying
\begin{align}
	\epsilon_2-\Omega
	&=
	\epsilon_1.
\end{align}
Consequently, the combination of Fermi distribution functions reduces to
\begin{align}
	\delta f
	&=
	f(\epsilon_1)
	+f(\epsilon_2-\Omega)
	-f(\epsilon_1+\Omega)
	-f(\epsilon_2)
	\notag \\
	&=
	f(\epsilon_1)
	+f(\epsilon_1)
	-f(\epsilon_2)
	-f(\epsilon_2)
	\notag \\
	&=
	2\Big(
	f(\epsilon_1)-f(\epsilon_2)
	\Big)
	\equiv
	2f_{12}.
\end{align}

\subsection{Geometric Representation of Nonlinear Electric Current}

For a better understanding of Eqs. (\ref{eq:j1_weak})--(\ref{eq:j3_weak}), we look into their connection to the geometric properties of electronic states. To this end, here we review geometric quantities that characterizes the electron bands. One such quantity is the Berry connection defined as $\mathcal{A}_{ij}^a \equiv -i \braket{u_i^s|\partial_{a} u_j^s}$, 
where $i,\; j$ denote the band indices.
For the interband transitions ($i \neq j$), the quantum metric and the interband Berry curvature are given by the product of the Berry connections \cite{ahn, ahn2022}:
\begin{align}
    \mathcal{G}_{ij}^{ab} \equiv {\rm Re} [\mathcal{A}_{ij}^a \mathcal{A}_{ji}^b],
    \quad 
    \mathcal{F}_{ij}^{ab} \equiv 2 {\rm Im} [\mathcal{A}_{ij}^a \mathcal{A}_{ji}^b].
\end{align}

Using the DC electric current formulas given in
Eqs.~(\ref{eq:j1_weak})--(\ref{eq:j3_weak}),
we obtain the explicit expressions
\begin{align}
    \bar J_{{\rm shift},  s}^{a;xy}
    = & 
    \pi (\Omega A_5)^2 \int \frac{d\bm{k}}{(2\pi)^3} R_{12, -}^{a;xy} 
    \notag \\
    \times &
    \left( \mathcal{G}_{12}^{xx}+\mathcal{G}_{12}^{yy} - \mathcal{F}_{12}^{xy} \right)
    f_{12} \delta(\epsilon_{1} - \epsilon_{2}+\Omega),\label{eq:j1}
    \\
    \bar J_{{\rm inj}, L ,s}^{a:xy}
    = & 
    - 2\pi\tau(\Omega A_5)^2
    \int\frac{d\bm{k}}{(2\pi)^3} \left(v_{11}^a-v_{22}^a\right)
    \notag \\
    \times &
    \left( \mathcal{G}_{12}^{xx}+\mathcal{G}_{12}^{yy} \right)
    f_{12}
    \delta(\epsilon_{1} - \epsilon_2 +\Omega),
    \\
    \bar J_{{\rm inj}, C, s}^{a;xy}
    = &  
    2\pi\tau(\Omega A_5)^2
    \int\frac{d\bm{k}}{(2\pi)^3} \left(v_{11}^a-v_{22}^a\right)
    \notag
    \\
    \times & \mathcal{F}_{12}^{xy}
    f_{12}
    \delta( \epsilon_{1} - \epsilon_2 +\Omega) .
    \label{eq:j3}
\end{align}
Here, the total injection and shift currents are defined as
$\bar J_{\rm inj}^{a;xy} \equiv \bar J_{\rm inj, L }^{a;xy} + \bar J_{\rm inj, C}^{a;xy}, \; \bar J_{\rm shift}^{a;xy} \equiv \bar J_{\rm shift, L }^{a;xy} + \bar J_{\rm shift, C}^{a;xy} $,
respectively, and 
$R_{12,\pm}^{a;bc} \equiv \partial_a \arg\left(v_{12}^b \pm i v_{12}^c\right)
+\mathcal A_{11}^a-\mathcal A_{22}^a$
is the shift vector.
The subscripts $L$ and $C$ denote the linear and circular components, respectively, associated with the symmetric and antisymmetric parts of the charge-current response kernel.
Detailed derivations of these expressions are provided in Appendix~\ref{app:geometric_current}.

For the Hamiltonian in Eq. (\ref{eq:j1})-(\ref{eq:j3}), the currents read
\begin{align}
    &\bar J_{{\rm shift},s}^{z;xy} = 0 ,\\
    &\bar J_{{\rm inj} , s}^{z;xy} = \bar J_{{\rm inj}, C, s}^{z;xy} =
    s \tau \frac{\Omega^2 A_5^2}{6\pi}\Theta
    \left(\Omega- 2\left|\mu + s\frac{\Delta}{2}\right|\right),
    \label{eq:inj}
\end{align}
in deriving which we used
\begin{align}
    f_{12} & =
    \Theta
    \left(
    \Omega
    +
    2\left(
    \mu + s\frac{\Delta}{2}
    \right)
    \right)-
    \Theta
    \left(
    \Omega
    -
    2\left(
    \mu - s\frac{\Delta}{2}
    \right)
    \right)
    \notag
    \\
    &=
    \Theta
    \left(
    \Omega
    -
    2\left|
    \mu + s\frac{\Delta}{2}
    \right|
    \right).
\end{align}
Here, we apply the absolute zero limit $f(\epsilon) \to \Theta (- \epsilon)$.
Consequently, the total DC current generated from a single Weyl node is given by
\begin{align}
    \bar J^{z;xy}_s 
    &\equiv
    \bar J_{{\rm shift}, s}^{z;xy}+\bar J_{{\rm inj} , s}^{z;xy}
    \notag
    \\
    &=
    s\tau
    \frac{\Omega^2 A_5^2}{6\pi}
    \Theta
    \left(
    \Omega
    -
    2\left|
    \mu + s\frac{\Delta}{2}
    \right|
    \right).
\end{align}

In a system without inversion symmetry where the chemical potentials of the two Weyl nodes differ, assuming that the $s = -1$ node has higher energy than $ s = + 1$ (see Fig. \ref{fig:weyl_nodes_position}), the total macroscopic current 
$\bar J_{\rm total}^z\equiv \bar J^z_{- } + \bar J^z_{+} $ exhibits piecewise behavior depending on the driving frequency $\Omega$:
\begin{align}
    \bar J_{\rm total}^z 
    &=
    \frac{\tau\Omega^2 A_5^2}{6\pi}
    \sum_{s=\pm1}
    s\,
    \Theta
    \left(
        \Omega
        -
        2\left|
        \mu + s\frac{\Delta}{2}
        \right|
    \right)
    \\
    &=
    \begin{cases}
        0 & (\Omega <  \Delta  + 2\mu ) \\
        \bar J^z_{+} & (\Delta  + 2\mu < \Omega <\Delta  - 2\mu) \\
        0 & (\Omega > \Delta  - 2\mu)
    \end{cases},
    \label{result}
\end{align}
If we assume the opposite configuration of the Weyl nodes, the “$+$” and “$-$” signs in (\ref{result}) are reversed.

\begin{figure*}
    \centering
    \includegraphics[width=0.9\linewidth]{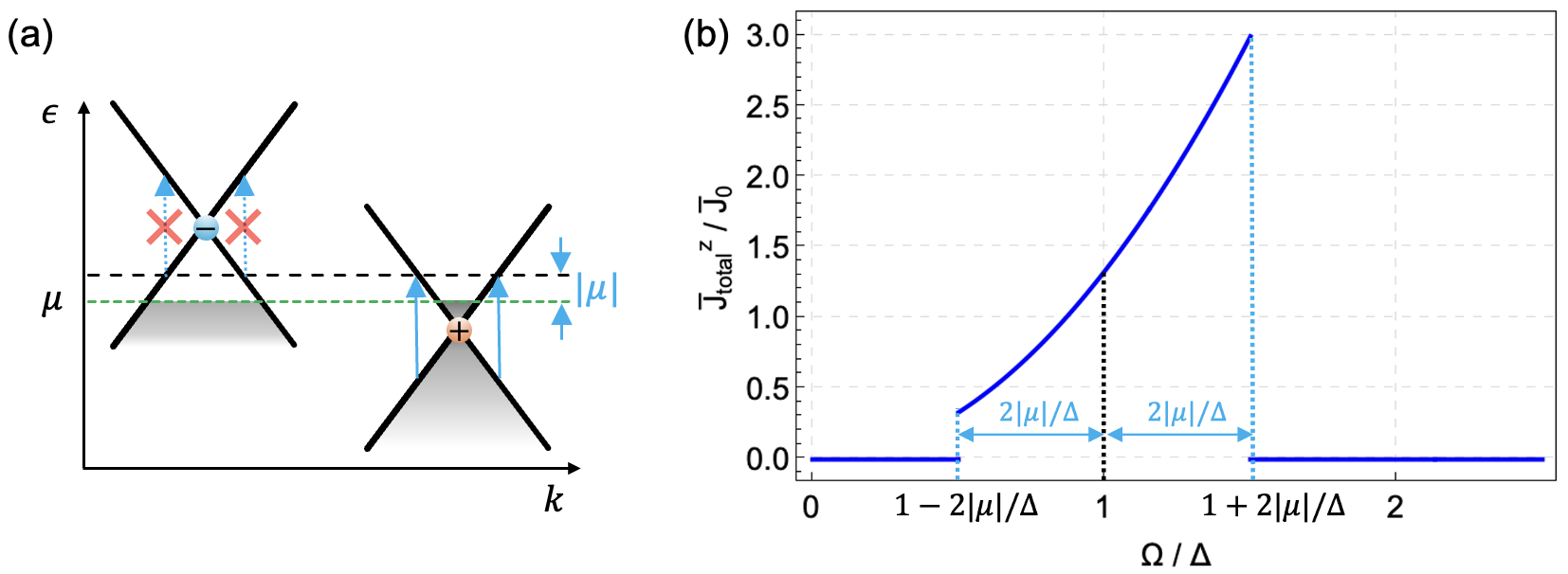}
    \caption{
    Frequency dependence of the current induced by spin precession for the potential $- \Delta /2 \leq \mu \leq 0$.
    (a) Schematic of the interband transitions induced by spin precession at two Weyl nodes separated in energy by $\Delta$.
    The gray regions represent occupied states, and the chemical potential lies a distance $|\mu|$ from the lower-energy $s = +1$ Weyl node.
    The blue arrows indicate drive-induced transitions. The transition at the $s = +1$ node is allowed, whereas that at the $s = -1$ node is forbidden by the emptiness of the initial state, as indicated by the red crosses.
    (b) Frequency dependence of the normalized total DC current
    $\bar{J}_{\mathrm{total}}/\bar{J}_0$ generated by spin precession, where
    $\bar{J}_0=\Delta A_5^2/(2\pi)$ and $\tau=1.0$.
    The spin-precession frequency $\Omega$ is normalized by the energy separation $\Delta$ between the Weyl nodes.
    For
    $\Omega/\Delta<1-2|\mu|/\Delta$,
    the transition at one Weyl node is energetically forbidden,
    while that at the other node is Pauli blocked;
    consequently, the total current vanishes.
    In the intermediate-frequency window $ 1-{2|\mu|}/{\Delta}<{\Omega}/{\Delta}<1+{2|\mu|}/{\Delta},$
    only one Weyl node contributes, resulting in a finite net current.
    For $\Omega/\Delta>1+2|\mu|/\Delta$,
    spin-precession-induced transitions are allowed at both Weyl nodes,
    and their oppositely directed current contributions cancel.
    }
    \label{fig:current_dependence}
\end{figure*}

\subsection{DC Current in a Tilted Weyl Semimetal}

In what follows, we analytically evaluate the spin-induced DC current in the tilted case. We will demonstrate that the current response exhibits discontinuous "jumps" in untilted Weyl semimetals, and further reveal that these abrupt jumps are significantly smoothed out when the Weyl cone is tilted. For simplicity in the subsequent analysis, we assume that the "$-$" node resides at a higher energy than the "$+$" node.

Compared to the original model, the Hamiltonian of the tilted Weyl semimetal includes an additional tilt term along the $z$-axis~\cite{ck}:
\begin{align}
    {\cal H}'_0 &= {\cal H}_0  - s u k_z \sigma_0,
\end{align}
where the tilt parameter $u$ is assumed to satisfy $u/v < 1$, which is the condition for a type-I Weyl semimetal. This modification shifts only the energy eigenvalues without altering the eigenstates:
\begin{align}
    \epsilon_1 ' &= \epsilon_1 - s uk_z , \quad \epsilon_2 ' = \epsilon_2 - s uk_z.
\end{align}
Consequently, the effect of the tilt enters the current integrals purely through the modified Fermi distribution function:
\begin{align}
    f_{12} \to f(\epsilon_1 - s u k_z) - f(\epsilon_2 - s u k_z).
\end{align}
We can explicitly determine the integration bounds governed by Fermi surface constraints. 
By the discussion, the DC electric current components are analytically evaluated as
\begin{align}
	\bar J_{{\rm inj},L,s}^{z;xy}
	&=
	\tau
	\frac{\Omega^2 A_5^2}{8\pi}
	\int_{-\bar k}^{\bar k}
	\frac{dk_z}{\bar k}
	\frac{k_z}{\bar k}
	\Big(
		1+\frac{k_z^2}{\bar k^2}
	\Big)
	f_{12}
    ,
	\\
	\bar J_{{\rm inj},C,s}^{z;xy}
	&=
	s\tau
	\frac{\Omega^2 A_5^2}{4\pi}
	\int_{-\bar k}^{\bar k}
	\frac{dk_z}{\bar k}
	\Big(
		\frac{k_z}{\bar k}
	\Big)^2
	f_{12} .
\end{align}
Here, we apply the absolute zero limit $f(\epsilon) \to \Theta ( - \epsilon)$ and we apply the following formulas to the theta function in the DC electric current:
\begin{align}
    f_{12} |_{s = +1}
    =&
    \Theta \Big[ \mu +s \frac{\Delta}{2} 
    +  v \bar k + s u k_z \Big]
    \notag \\
    &
    - \Theta \Big[\mu +s \frac{\Delta}{2} 
    -  v \bar k + s u k_z \Big]
    .
\end{align}
where the resonance condition $v\bar k=\Omega/2$ has been used.

On the resonant surface $k=\bar k$, the relevant threshold momenta
are determined by the Fermi-level crossings of the conduction band
at the $s=+1$ node and the valence band at the $s=-1$ node:
\begin{align}
    \epsilon_2^{\prime +}(K_+)
    &=
    v\bar k
    -
    \Big(
        \mu+\frac{\Delta}{2}
    \Big)
    -
    uK_+
    =
    0,
    \notag
    \\
    \epsilon_1^{\prime -}(K_-)
    &=
    -v\bar k
    -
    \Big(
        \mu-\frac{\Delta}{2}
    \Big)
    +
    uK_-
    =
    0.
\end{align}
Using $v\bar k=\Omega/2$, the two threshold momenta are compactly
written as
\begin{align}
    K_\pm
    &=
    \frac{v\bar k}{u}
    \Big(
        1-\frac{\Delta\pm2\mu}{2 v \bar k}
    \Big).
\end{align}

After combining the contributions from the two Weyl nodes,
the Pauli-blocking constraints can be expressed in terms of
$K_+$ and $K_-$.
Taking into account the physical range $-\bar k\leq k_z\leq \bar k$,
the relevant integration interval is determined by the intersection
of these threshold momenta with the resonant momentum sphere.

Considering the regime in which the $s=-1$ Weyl node is located at
$\Delta/2-\mu>0$, while the $s=+1$ Weyl node is located at
$-\Delta/2-\mu<0$, we obtain
\begin{align}
    & f_{12}|_{s = + 1} -  f_{12}|_{s = - 1}
    \notag \\
    & =
    1 - \Theta\Big[
    k_z - \frac{v}{u} \bar k \Big( 1 - \frac{\Delta + 2\mu}{ 2 v \bar k} \Big)
    \Big]
    \notag \\
    & \quad 
    -  \Big\{ \Theta
    \Big[
    \frac{v}{u} \bar k \Big( 1- \frac{\Delta - 2\mu}{2v \bar k}  \Big) - k_z
    \Big] - 0 \Big\}
    \notag \\
    & =
    \Theta (  K_+ -  k_z  ) - \Theta ( K_- - k_z  )
    ,
    \\
    & f_{12}|_{s = + 1} +  f_{12}|_{s = - 1}
    \notag \\
    & =
    1 - \Theta\Big[
    k_z - \frac{v}{u} \bar k \Big( 1 - \frac{\Delta + 2\mu}{ 2 v \bar k} \Big)
    \Big]
    \notag \\
    & \quad 
    + \Theta
    \Big[
    \frac{v}{u} \bar k \Big( 1- \frac{\Delta - 2\mu}{2v \bar k}  \Big) - k_z
    \Big] - 0
    \notag \\
    & =
    \Theta (  K_+ -  k_z  )
    +
    \Theta (K_- -  k_z )
    \notag \\
    & =
    2 \Theta (  K_+ -  k_z  )
    -
    \Big( 
    \Theta (  K_+ -  k_z  ) - \Theta ( K_- - k_z  )
    \Big)
    .
\end{align}
Consequently,
\begin{align}
	&
	\bar J_{{\rm inj},C,+}^{z;xy}
	+
	\bar J_{{\rm inj},C,-}^{z;xy}
	\notag
	\\
	={}&
	\tau
	\frac{\Omega^2 A_5^2}{4\pi}
	\int_{\max\{-\bar k,K_-\}}^{\min\{\bar k,K_+\}}
	\frac{dk_z}{\bar k}
	\Big(
		\frac{k_z}{\bar k}
	\Big)^2,
	\notag
	\\
	&
	\bar J_{{\rm inj},L,+}^{z;xy}
	+
	\bar J_{{\rm inj},L,-}^{z;xy}
	\notag
	\\
	={}&
	2\tau
	\frac{\Omega^2 A_5^2}{8\pi}
	\int_{-\bar k}^{\min \{\bar k,K_+ \}}
	\frac{dk_z}{\bar k}
	\frac{k_z}{\bar k}
	\Big(
		1+\frac{k_z^2}{k^2}
	\Big)
	\notag
	\\
	&
	-
	\tau
	\frac{\Omega^2 A_5^2}{8\pi}
	\int_{\max\{-\bar k,K_-\}}^{\min\{\bar k,K_+\}}
	\frac{dk_z}{\bar k}
	\frac{k_z}{\bar k}
	\Big(
		1+\frac{k_z^2}{k^2}
	\Big).
\end{align}

\begin{figure}[tbp]
    \centering
    \includegraphics[width=1\linewidth]{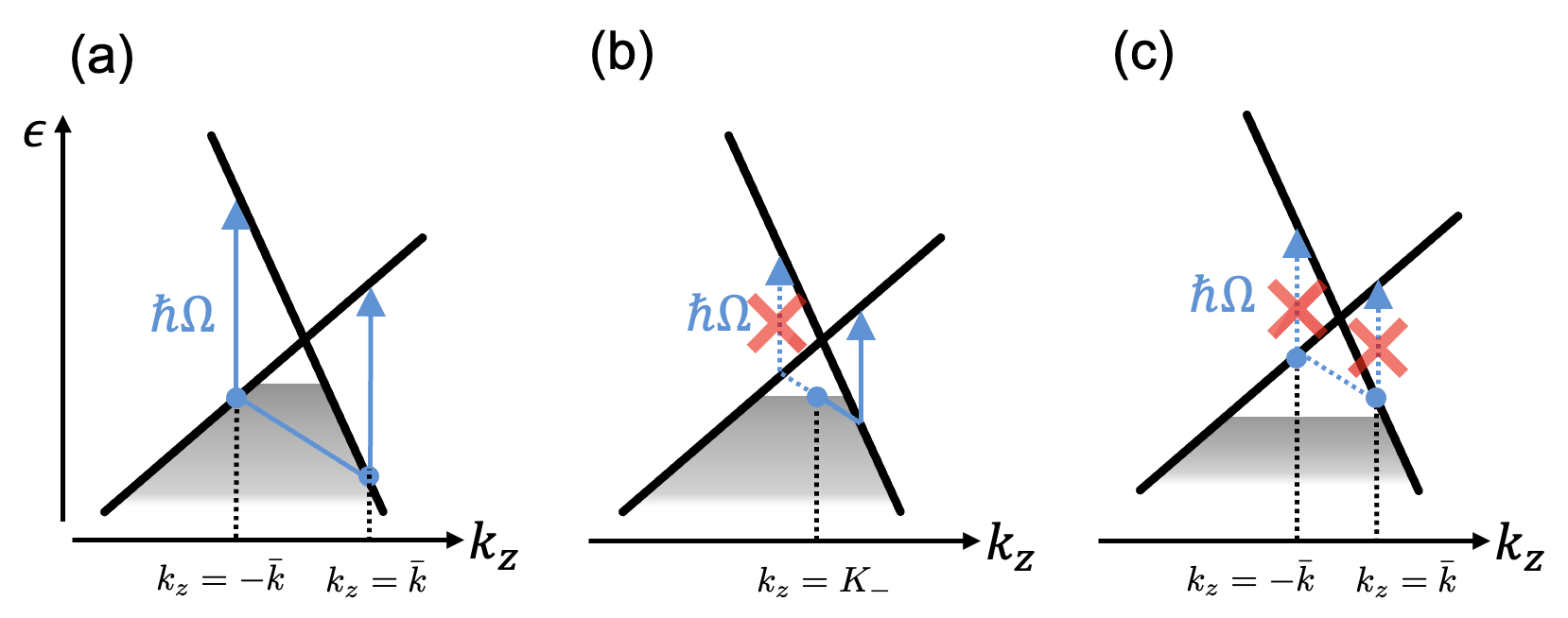}
    \caption{Schematic illustrating the different cases for the lower integration bound. (a) When the resonant surface lies entirely below the chemical potential, the integration range begins at $-\bar k = \max \{ -\bar k, K_-\}$. (b) When the chemical potential intersects the resonant surface, only the portion below the chemical potential contributes, restricting the lower integration bound to $K _- = \max \{ -\bar k, K_-\}$. (c) When $K_- > \bar k$, the integration range completely collapses.}
    \label{fig:tilt}
\end{figure}

By comparing the integration bounds under the type-I condition
$u/v<1$, the explicit form of the integral can be determined
for each parameter regime.
In the untilted limit, $u\to0$, the Fermi-surface constraint
reduces to that of the untilted Weyl semimetal, reproducing the
results obtained in the previous section.
The integration range
$k_z\in[\max\{-\bar k,K_-\},\min\{\bar k,K_+\}]$
is determined by the intersection between the Pauli-allowed region
and the resonant momentum surface (see Fig. \ref{fig:tilt}) \cite{Mukherjee2017}.

By comparing the integration bounds under the type-I condition
$0<u/v<1$, the integral can be evaluated explicitly in each
parameter regime.
In the untilted limit, $u\to0$, the constraints imposed by the
occupation factors reduce to those of an untilted Weyl semimetal,
and the results obtained in the previous section are recovered.

The integration interval
$k_z\in[\max\{-\bar k,K_-\},\min\{\bar k,K_+\}]$
is determined by the overlap between the constant-energy shell
selected by the resonance condition and the region in which the
interband transitions are not Pauli blocked
(see Fig. \ref{fig:tilt}) \cite{Mukherjee2017}.
For the isotropic Weyl dispersion considered here, the resonance
condition fixes the momentum magnitude to
$\bar k=\Omega/(2v)$.

For $\mu/\Delta\leq0$ and $\Omega/\Delta>1$, the integration bounds
can be classified as follows:
\begin{align}
	&\text{Case I}:\quad
	1
	<
	\frac{\Omega}{\Delta}
	<
	\frac{1-2\mu/\Delta}{1+u/v},
	\quad 
	\frac{u}{v}
	<
	-\frac{2\mu}{\Delta},
	\\
	&\text{Case II}:\quad
	\frac{\Omega}{\Delta}
	>
	\max\Bigg\{
		1,
		\frac{1-2\mu/\Delta}{1+u/v},
		\frac{1+2\mu/\Delta}{1-u/v}
	\Bigg\},
	\\
	&\text{Case III}:\quad
	1
	<
	\frac{\Omega}{\Delta}
	<
	\frac{1+2\mu/\Delta}{1-u/v},
	\quad
	\frac{u}{v}
	>
	-\frac{2\mu}{\Delta}.
\end{align}
Case I therefore occurs only when
$u/v<-2\mu/\Delta$, whereas Case III occurs only when
$u/v>-2\mu/\Delta$.
At $u/v=-2\mu/\Delta$, the two frequency thresholds coincide with
$\Omega/\Delta=1$, and only Case II is realized within the assumed
frequency range.

For the integration interval to have a finite width, the two
conditions
\begin{align}
	K_-&<\bar k,
	\quad 
	K_+ >-\bar k
\end{align}
must be satisfied.
Solving these inequalities for $\Omega/\Delta$ gives
\begin{align}
	\frac{\Omega}{\Delta}
	&<
	\frac{1-2\mu/\Delta}{1-u/v},
	\quad 
	\frac{\Omega}{\Delta}
	>
	\frac{1+2\mu/\Delta}{1+u/v},
\end{align}
respectively.
For $\mu/\Delta\leq0$ and $0<u/v<1$, the second threshold is
smaller than unity and is therefore automatically satisfied under
the assumption $\Omega/\Delta>1$.
Consequently, the condition for a nonvanishing integration interval
reduces to
\begin{align}
	1
	<
	\frac{\Omega}{\Delta}
	<
	\frac{1-2\mu/\Delta}{1-u/v}.
\end{align}
In the untilted limit, this frequency window becomes
$1<\Omega/\Delta<1-2\mu/\Delta$, consistently reproducing the
result for the untilted Weyl semimetal.

\begin{table}[t]
\centering
\caption{
Nonvanishing current components and their azimuthally integrated
geometric kernels for an isotropic Weyl cone with a tilt along the
$z$ direction.
The expressions are shown for $\eta=+1$ and the reference chirality
$s=+1$.
All current-axis components not listed in the table vanish upon integration over the azimuthal angle $\phi$.
}
\label{tab:z-tilt-current-components}
\begin{tabular}{ccc}
\toprule
Precession
&
Response
&
Nonzero current component
\\
\hline
\hline

$xy$
&
Linear injection
&
$\bar{J}_{\mathrm{inj,L}}^{z;xy}$

\\

$xy$
&
Circular injection
&
$\bar{J}_{\mathrm{inj,C}}^{z;xy}$
\\

\midrule

$yz$
&
Linear injection
&
$\bar{J}_{\mathrm{inj,L}}^{z;yz}$

\\

$yz$
&
Circular injection
&
$\bar{J}_{\mathrm{inj,C}}^{x;yz}$
\\

$yz$
&
Shift
&
$\bar{J}_{\mathrm{shift}}^{y;yz}$
\\

\midrule

$zx$
&
Linear injection
&
$\bar{J}_{\mathrm{inj,L}}^{z;zx}$
\\

$zx$
&
Circular injection
&
$\bar{J}_{\mathrm{inj,C}}^{y;zx}$
\\

$zx$
&
Shift
&
$\bar{J}_{\mathrm{shift}}^{x;zx}$
\\

\bottomrule
\end{tabular}
\end{table}

\begin{figure*}[tbp]
    \centering
    \includegraphics[width=1 \linewidth]{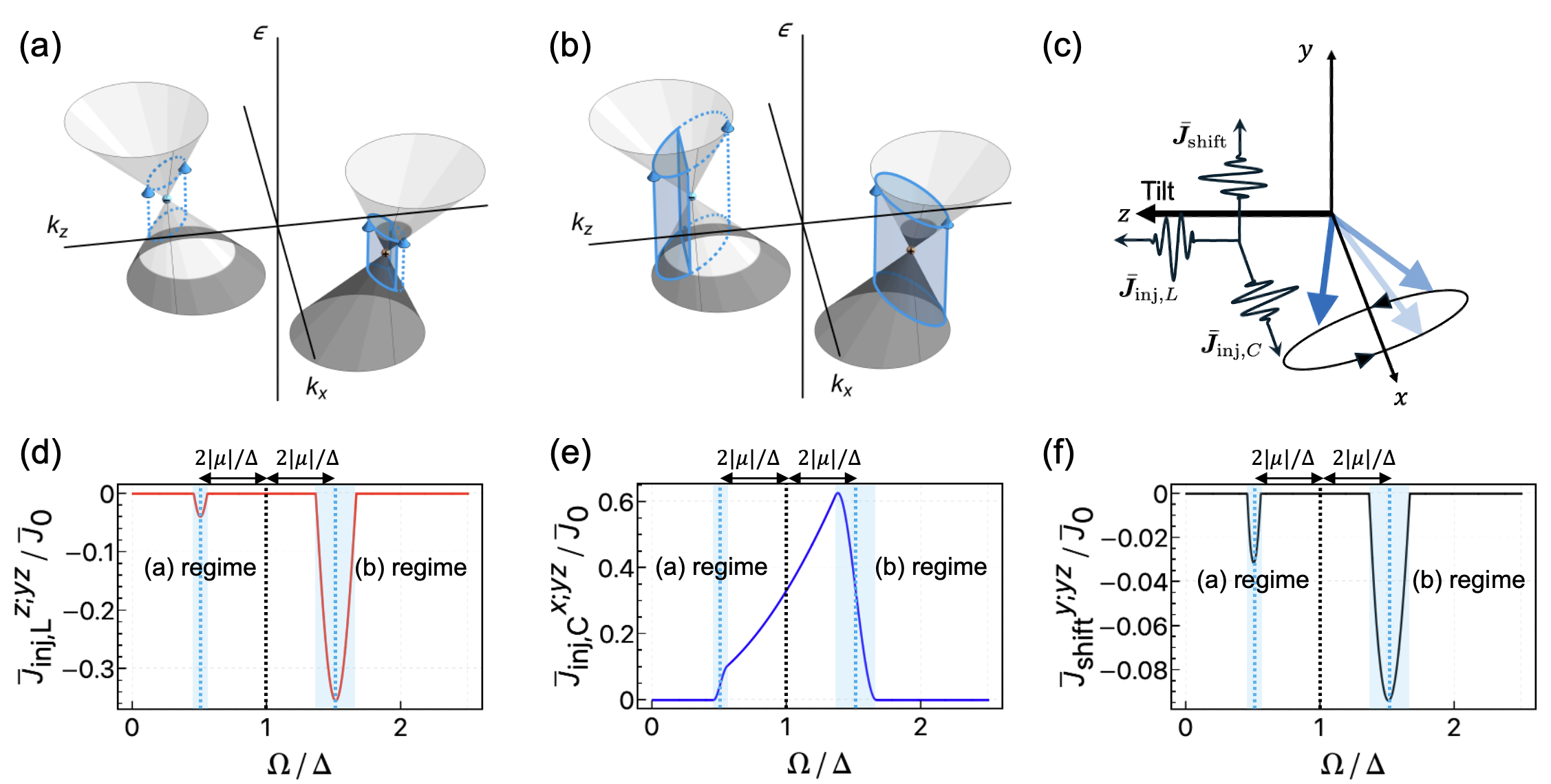}
    \caption{ Frequency dependence and directions of the dc currents induced by magnetization precession in the $yz$ plane for a Weyl system with a
    $z$-directed tilt.
    (a) and (b) Schematics of the resonant interband transitions at the
    two Weyl nodes in the lower-frequency onset regime and the
    higher-frequency cancellation regime, respectively.
    The solid blue arrows represent Pauli-allowed transitions, whereas
    the dotted blue contours indicate the portions of the resonant
    surfaces excluded by Pauli blocking.
    The finite tilt makes the allowed transition phase space depend on
    $k_z$.
    (c) Relative directions of the tilt, magnetization precession, and
    the induced currents.
    The linear injection current is parallel to the tilt,
    $\bm{\bar J}_{{\rm inj},L}\parallel\bm t$; the circular injection
    current is parallel to the precession-helicity vector,
    $\bm{\bar J}_{{\rm inj},C}\parallel\bm H$; and the shift current is
    transverse to both,
    $\bm{\bar J}_{\rm shift}\parallel\bm t\times\bm H$, where
    $\bm H=i\bm A_{5,\Omega}\times\bm A_{5,\Omega}^{*}$.
    For the geometry shown here, these relations give
    $\bar J_{{\rm inj},L}^{z;yz}$,
    $\bar J_{{\rm inj},C}^{x;yz}$, and
    $\bar J_{\rm shift}^{y;yz}$.
    (d)--(f) Frequency dependences of the normalized linear injection
    current $\bar J_{{\rm inj},L}^{z;yz}/\bar J_0$, circular injection
    current $\bar J_{{\rm inj},C}^{x;yz}/\bar J_0$, and shift current
    $\bar J_{\rm shift}^{y;yz}/\bar J_0$, respectively, where
    $\bar{J}_0=\Delta A_5^2/(2\pi)$.
    The blue dotted vertical lines mark
    $\Omega/\Delta=1-2|\mu|/\Delta$ and
    $\Omega/\Delta=1+2|\mu|/\Delta$, while the black dotted line marks
    $\Omega/\Delta=1$. We set $\tau\Delta=1.0$.
    The horizontal arrows indicate the separation
    $2|\mu|/\Delta$ between these characteristic frequencies.
    The light-blue shaded regions denote the frequency windows broadened
    by the finite Weyl-cone tilt and correspond to the regimes illustrated in panels (a) and (b).
    In the present geometry, the linear shift current vanishes, so that
    the shift current shown in panel (f) is entirely helicity odd.
}
    \label{fig:placeholder}
\end{figure*}

\subsection{The case of spin precession in the another plane}

In this subsection, we investigate the current response when the spin precession occurs in the $bc$-plane, where $b, c \in \{x, y, z\}$. For instance, when the precession is confined to the $yz$-plane , i.e., $(b, c) = (y, z)$, the magnetization vector is given by $\bm{m}(t) = \big( \sqrt{1-m_\perp^2}, m_\perp \cos \Omega t, m_\perp \sin \Omega t \big)^T$. Correspondingly, the relevant subspace of the effective Floquet Hamiltonian takes the form
\begin{equation}
    {H'}_{F, bc}^{2\times2} =
    \begin{pmatrix}
        \epsilon_1' + |\Omega| & -s\frac{A_5}{\sqrt{2}} (v_{12}^b - i v_{12}^c) \\
        -s\frac{A_5}{\sqrt{2}} (v_{21}^b + i v_{21}^c) & \epsilon_2' 
    \end{pmatrix}.
\end{equation}
The velocity operator along the $a$-direction is obtained by taking the momentum derivative of the Hamiltonian:
\begin{align}
    &\tilde{v}'^{a;bc} =
    \notag \\
    &
    \begin{pmatrix}
        \partial_a \epsilon_1' & 
        -s \frac{A_5}{\sqrt{2}} 
        ((\partial_a  v^b)_{12} - i (\partial_a  v^c)_{12}) \\
        -s \frac{A_5}{\sqrt{2}} 
        ((\partial_a  v^b)_{21} + i (\partial_a  v^c)_{21}) & \partial_a \epsilon_2'
    \end{pmatrix}.
\end{align}
Following the same procedure as in Eqs.~\eqref{eq:j1}--\eqref{eq:j3}, the DC current induced by magnetization precession in the $bc$ plane can be obtained by applying the cyclic permutation
$(x,y,z)\to(a,b,c)$ to the full Hamiltonian, including the tilt term.
Under this transformation, both $\bar J_{\rm shift}^{a;bc}$ and $\bar J_{\rm inj}^{a;bc}$ can be nonzero.

We now consider the dc current induced by magnetization precession
in the $yz$ plane for a Weyl system with a $z$-directed tilt.
Here, each current denotes the sum of the contributions from the two
Weyl nodes with opposite chiralities.
After integration over the azimuthal angle $\phi$, the potentially nonzero components are the shift current along the $y$ direction, the circular injection current along the $x$ direction, and the linear injection current along the $z$ direction:
\begin{align}
    \bar J_{\rm shift}^{y;yz}
    &=
    2\cdot \frac{\Omega A_5^2}{16\pi}
    \int_{-\bar k}^{\min\{\bar k,K_+\}}
    \frac{dk_z}{\bar k}
    \frac{k_z}{\bar k}
    \notag\\
    & \quad 
    - \frac{\Omega A_5^2}{16\pi}
    \int_{\max\{-\bar k,K_-\}}^{\min\{\bar k,K_+\}}
    \frac{dk_z}{\bar k}
    \frac{k_z}{\bar k}
    ,
    \label{eq:shift-y-yz}
    \\
    \bar J_{{\rm inj},C}^{x;yz}
    &=
    \tau\frac{\Omega^2 A_5^2}{8\pi}
    \int_{\max\{-\bar k,K_-\}}^{\min\{\bar k,K_+\}}
    \frac{dk_z}{\bar k}
    \Big[
        1-\left(\frac{k_z}{\bar k}\right)^2
    \Big],
    \label{eq:circular-injection-x-yz}
    \\
    \bar J_{{\rm inj},L}^{z;yz}
    &=
    2\tau\frac{\Omega^2 A_5^2}{16\pi}
    \int_{-\bar k}^{\min\{\bar k,K_+\}}
    \frac{dk_z}{\bar k}
    \frac{k_z}{\bar k}
    \Big[
        3-\left(\frac{k_z}{\bar k}\right)^2
    \Big]
    \notag\\
    &
    - \tau\frac{\Omega^2 A_5^2}{16\pi}
    \int_{\max\{-\bar k,K_-\}}^{\min\{\bar k,K_+\}}
    \frac{dk_z}{\bar k}
    \frac{k_z}{\bar k}
    \Big[
        3-\left(\frac{k_z}{\bar k}\right)^2
    \Big].
    \label{eq:linear-injection-z-yz}
\end{align}
The integrals are understood to vanish when their lower boundary is
greater than or equal to their upper boundary (see Fig.~\ref{fig:placeholder}).

The directions of these current components can be understood as follows.
The linear injection current is even under reversal of the magnetization-precession direction.
For the isotropic Weyl model with a $z$-directed tilt, azimuthal integration eliminates the $x$ and $y$ components, leaving only the component parallel to the tilt direction,
$\bm{\bar J}_{{\rm inj},L}\parallel\bm t.$
Since the electric current is a polar vector while the symmetric quadratic combination of the driving fields is inversion even, a nonzero net linear injection current requires inequivalent contributions from inversion-related Weyl nodes.
In an inversion-symmetric system, their contributions cancel.

The circular injection current is odd under reversal of the magnetization-precession direction and flows along the spin-helicity vector
$\bm H \equiv i\bm A_{5,\Omega}\times\bm A_{5,\Omega}^{*}.$
For precession in the $yz$ plane, $\bm H$ points along the $x$ direction.
More precisely, the antisymmetric part of the driving-field bilinear couples to the Berry curvature, producing a circular injection current along $\bm H$.
Because the Berry curvature changes sign between Weyl nodes of opposite chiralities, their circular injection currents are oppositely directed.
A finite total current therefore requires unequal transition phase spaces at the two Weyl nodes.

The shift current is a polar vector and is odd under spatial inversion.
The helicity $\bm H$ is an axial vector, whereas the tilt $\bm t$ is a polar vector.
Therefore, $\bm t\times\bm H$ transforms as a polar vector:
\begin{equation}
    \mathcal P:
    \bm t\mapsto-\bm t,
    \;
    \bm H\mapsto\bm H,
    \;
    \bm t\times\bm H
    \mapsto-\bm t\times\bm H.
\end{equation}
The symmetry-allowed direction of the shift current is therefore
$\bm{\bar J}_{\rm shift} \parallel \bm t\times\bm H$.
For a $z$-directed tilt and $yz$-plane precession, $\bm t\parallel\hat{\bm z}$ and $\bm H\parallel\hat{\bm x}$, so that $\bm{\bar J}_{\rm shift}\parallel\hat{\bm y}$, up to the sign of the microscopic response coefficient \cite{wy2021, ahn}.


\subsection{Direction of the Current Response}

In the preceding analysis, we focused on a fixed direction of the
magnetization precession.
To discuss the reversal of the precession direction, we separate the
magnitude and direction of the frequency by introducing
$\eta\equiv{\rm sgn}(\Omega)$ and $\Omega=\eta|\Omega|$.
For the magnetization precession in the $yz$ plane, we write
$\bm m(t)=\left(\sqrt{1-m_\perp^2},m_\perp\cos|\Omega|t,
\eta m_\perp\sin|\Omega|t\right)^T$.
With this convention, the resonant Floquet subspace is always
constructed from the one-photon-dressed valence band and the bare
conduction band, irrespective of the sign of $\Omega$.
The corresponding effective Floquet Hamiltonian is
\begin{align}
	H_{F,\eta}^{2\times2}
	&=
	\begin{pmatrix}
		\epsilon_1+|\Omega|
		&
		-s\frac{A_5}{\sqrt{2}}
		\left(
			v_{12}^y-i\eta v_{12}^z
		\right)
		\\
		-s\frac{A_5}{\sqrt{2}}
		\left(
			v_{21}^y+i\eta v_{21}^z
		\right)
		&
		\epsilon_2
	\end{pmatrix}.
\end{align}

We define
\begin{align}
	R_{12,\eta}^{a;yz}
	&\equiv
	\partial_a{\rm arg}
	\left(
		v_{12}^y+i\eta v_{12}^z
	\right)
	+
	\mathcal A_{11}^a
	-
	\mathcal A_{22}^a.
\end{align}
It then follows that $R_{12,-\eta}^{a;zy}=R_{12,\eta}^{a;yz}. $
Note that the off-diagonal Floquet matrix element for a given
precession helicity $\eta$ is proportional to
$v_{12}^y-i\eta v_{12}^z$.
According to the above definition, the corresponding shift vector is
therefore $R_{12,-\eta}^{a;yz}$.

The reversal of the precession direction is related to the symmetric and antisymmetric decomposition with respect to the two driving indices.
The symmetric and antisymmetric parts of either the injection or shift current can equivalently be obtained from the currents for the two opposite precession directions:
\begin{align}
	\bar J_{X,L}^{a;yz}
	& = 
	\frac{\bar J_{X,+}^{a;yz}+\bar J_{X,-}^{a;yz}}{2},
	\; 
	\bar J_{X,C}^{a;yz}
	=
	\frac{\bar J_{X,+}^{a;yz}-\bar J_{X,-}^{a;yz}}{2}
	,
\end{align}
where $X={\rm inj,shift}$ and the subscripts $\pm$ denote $\eta=\pm1$, respectively.

For the injection current, this decomposition is particularly
transparent in terms of the quantum geometric tensor.
The quantum-metric contribution is symmetric under the interchange of
the two driving indices, whereas the Berry-curvature contribution is
antisymmetric.
Accordingly,
\begin{align}
	\left(
		v_{11}^a-v_{22}^a
	\right)
	\left(
		\mathcal G_{12}^{yy}
		+
		\mathcal G_{12}^{zz}
	\right)
	&\to
	\left(
		v_{11}^a-v_{22}^a
	\right)
	\left(
		\mathcal G_{12}^{yy}
		+
		\mathcal G_{12}^{zz}
	\right),
	\\
	\left(
		v_{11}^a-v_{22}^a
	\right)
	\mathcal F_{12}^{yz}
	&\to
	\eta
	\left(
		v_{11}^a-v_{22}^a
	\right)
	\mathcal F_{12}^{yz}.
\end{align}
Thus, the linear injection current is even under
$\eta\to-\eta$, while the circular injection current is odd:
\begin{align}
	\bar J_{\rm inj,L}^{z;yz}(-\eta)
	&=
	\bar J_{\rm inj,L}^{z;yz}(\eta),
	\quad 
	\bar J_{\rm inj,C}^{z;yz}(-\eta)
	=
	-\bar J_{\rm inj,C}^{z;yz}(\eta).
\end{align}

For the shift current, however, the precession dependence cannot be
determined from the quantum metric and Berry curvature alone, because
the shift vector itself depends on the precession helicity.
The shift vector associated with the Floquet matrix element
$v_{12}^y-i\eta v_{12}^z$ is $R_{12,-\eta}^{a;yz}$.
Therefore, the relevant helicity-dependent combination is
\begin{align}
	\left|
		v_{12}^y-i v_{12}^z
	\right|^2
	R_{12,-}^{a;yz}
	&\to
	\left|
		v_{12}^y-i\eta v_{12}^z
	\right|^2
	R_{12,-\eta}^{a;yz}.
	\label{eq:eta_shift_full}
\end{align}
Although the velocity matrix element can be written as
\begin{align}
	\left|
		v_{12}^y-i\eta v_{12}^z
	\right|^2
	&=
	(\epsilon_{12})^2
	\left(
		\mathcal G_{12}^{yy}
		+
		\mathcal G_{12}^{zz}
		-
		\eta\mathcal F_{12}^{yz}
	\right),
\end{align}
the factors proportional to the quantum metric and Berry curvature
cannot be separately identified with the linear and circular shift
currents before the $\eta$ dependence of
$R_{12,-\eta}^{a;yz}$ is included.
For the present Weyl Hamiltonian, the direct evaluation of Eq.~\eqref{eq:eta_shift_full} gives, after the azimuthal integration,
\begin{align}
	\bar J_{{\rm shift},-}^{y;yz}
	&=
	-\bar J_{{\rm shift},+}^{y;yz}.
\end{align}
Consequently,
\begin{align}
	\bar J_{\rm shift,L}^{y;yz}
	&=
	0,
	\quad
	\bar J_{\rm shift,C}^{y;yz}
	=
	\bar J_{{\rm shift},+}^{y;yz} = - \bar J_{{\rm shift},-}^{y;yz}.
\end{align}
Therefore, the finite shift current obtained for the $yz$-plane
precession is purely antisymmetric with respect to the two driving
indices and reverses its direction when the sense of magnetization precession is reversed.

Note that the operation changing sign of $\eta$ correspond exchanging subscripts of input.

\section{Numerical calculation}

To estimate the magnitude of these current density in a realistic MWSM accessible to experiment, we consider the compensated ferrimagnetic inverse-Heusler compound \ce{Ti2MnAl}, which has been predicted to host Weyl nodes near the Fermi energy~\cite{Shi-Muechler-etal2018}.
Its compensated ferrimagnetic order breaks $\mathcal{T}$ symmetry, and its noncentrosymmetric crystal structure breaks $\mathcal{P}$ symmetry.
These band-structure and symmetry properties motivate the application of the continuum-model analysis.

An effective tight-binding study of \ce{Ti2MnAl} indicates strong exchange coupling between the Ti and Mn sublattices, together with weak magnetic anisotropy of the ferrimagnetic order~\cite{Meguro2024,meguro2025topological,furusho2025fermi}.
These results motivate a quantitative investigation of spin precession driven by an oscillating magnetic field.
In the following, we solve the coupled Landau--Lifshitz--Gilbert equations to determine the resonance frequency and precession amplitude, which we then use to estimate the induced current density.

\subsection{Landau--Lifshitz--Gilbert Equation of Ti${}_2$MnAl }
We use the $4\times4$ model of Ref.~\cite{FurushoP}, with unit magnetization vectors satisfying $|\bm m ^a|=1$. At Weyl nodes $\bm k = \bm k _W$, the Hamiltonian is
\begin{align}
    &H_{\rm TMA}(\bm k  ; \bm m ^{\rm Ti1}, \bm m ^{\rm Mn})
    \notag \\
    &=
    \begin{pmatrix}
        \xi_{\rm Ti1} I_2 + \bm b_{\rm Ti1} \cdot \bm \sigma &
        (d_1 - id_2)I_2 \\
        (d_1 + id_2)I_2 & \xi_{\rm Mn} I_2 + \bm b_{\rm Mn} \cdot \bm \sigma
    \end{pmatrix},
\end{align}
where
\begin{align}
    &\xi_a(\bm k)
    = \epsilon_a - 2t_{aa}f(\bm k),
    \quad a = {\rm Ti1,\,Mn}, \\
    &\bm b_{\rm Ti1}(\bm k)
    = -J_{\rm exc}^{\rm Ti1}\bm m^{\rm Ti1}
      -2\lambda_{\rm Ti1}\bm n_{\rm KM}(\bm k), \\
    &\bm b_{\rm Mn}(\bm k)
    = -J_{\rm exc}^{\rm Mn}\bm m^{\rm Mn}
      +2\lambda_{\rm Mn}\bm n_{\rm KM}(\bm k), \\
    &d_1(\bm k)
    = -2t_{\mathrm{Mn,Ti1}}
       \sum_{i=x,y,z}\cos\frac{k_i}{2}
       \notag \\
    &\quad -4\bigl(
       t_{\mathrm{Ti1,Ti2,Mn}}
       +t_{\mathrm{Ti1,Al,Mn}}
       \bigr)
       \prod_{i=x,y,z}\cos\frac{k_i}{2}, \\
    &d_2(\bm k)
    = 4\bigl(
       t_{\mathrm{Ti1,Ti2,Mn}}
       -t_{\mathrm{Ti1,Al,Mn}}
       \bigr)
       \prod_{i=x,y,z}\sin\frac{k_i}{2}.
\end{align}
Here, we use the parameter values~\cite{FurushoP}:
\begin{align}
    \epsilon_{\rm Ti1}
    &= 1.1074\;{\rm eV},
    \quad \epsilon_{\rm Mn}=-0.5066\;{\rm eV},
    \notag \\
    J_{\rm exc}^{\rm Ti1}
    &= 1.6154\;{\rm eV},
    \quad J_{\rm exc}^{\rm Mn}=0.7836\;{\rm eV},
    \notag \\
    \lambda_{\rm Ti1}
    &= \lambda_{\rm Mn}=0.015\;{\rm eV},
    \notag \\
    t_{\mathrm{Ti1,Ti1}}
    &= 0.1768\;{\rm eV},
    \quad t_{\mathrm{Mn,Mn}}=-0.102\;{\rm eV},
    \notag \\
    t_{\mathrm{Mn,Ti1}}
    &= -0.0544\;{\rm eV},
    \notag \\
    t_{\mathrm{Ti1,Ti2,Mn}}
    &= 0.076\;{\rm eV},
    \quad t_{\mathrm{Ti1,Al,Mn}}=0.05\;{\rm eV}.
    \label{eq:parameters}
\end{align}
The dimensionless hopping and spin-orbit coupling (SOC) structure factors are
\begin{align}
    f (\bm k) & = 2\biggl(
    \cos\frac{k_x}{2}\cos\frac{k_y}{2}
    +\cos\frac{k_y}{2}\cos\frac{k_z}{2}
    +\cos\frac{k_z}{2}\cos\frac{k_x}{2}\biggr), \\
    &\bm n _{\rm KM} (\bm k) = -\sqrt{2}
\begin{pmatrix}
    \sin\frac{k_x}{2}\left(\cos\frac{k_y}{2}-\cos\frac{k_z}{2}\right) \\
    \sin\frac{k_y}{2}\left(\cos\frac{k_z}{2}-\cos\frac{k_x}{2}\right) \\
    \sin\frac{k_z}{2}\left(\cos\frac{k_x}{2}-\cos\frac{k_y}{2}\right)
\end{pmatrix}.
\end{align}
The components of $\bm k$ refer to the cubic crystal axes, with the
lattice constant set to unity.
The four band energies, labeled by $\nu=1,\ldots,4$, satisfy
\begin{align}
    \det[ E_{\nu} (\bm k;\bm m ^{\rm Ti1}, \bm m ^{\rm Mn} ) I_4 
    - H_{\rm TMA}(\bm k ; \bm m ^{\rm Ti1}, \bm m ^{\rm Mn} ) ] = 0.
\end{align}
Their grand canonical potential per primitive cell is
\begin{align}
    &\Phi_{\rm el}(\bm m ^{\rm Ti1}, \bm m ^{\rm Mn})
    \notag \\
    &\equiv - \frac{k_{\rm B} T}{N_k}\sum_{\bm k , \nu}
    \log (1 + e^{ - [E_{\nu} (\bm k;\bm m ^{\rm Ti1}, \bm m ^{\rm Mn} ) - \mu ]\beta }),
\end{align}
where $\beta$ is the inverse temperature $(k_{\rm B}T)^{-1}$ and $N_k$ is the number of sampled wave vectors. The electron number within this model is
\begin{align}
    n_e \equiv \frac{1}{N_k}\sum_{\bm k , \nu}
    \frac{1}{1+e^{[E_{\nu} (\bm k;\bm m ^{\rm Ti1}, \bm m ^{\rm Mn} )-\mu]\beta}}.
\end{align}
The electronic free energy is
\begin{align}
    F_{\rm el } &\equiv
    \Phi_{\rm el} + \mu n_e
    \notag \\
    &= \Phi_0 + \frac{1}{2} \delta \bm m ^T C \delta \bm m
    + \mathcal{O} (|\delta \bm m|^3) + \mu n_e,
\end{align}
where $\delta \bm m = (0,  m ^{\rm Ti1}_y , m ^{\rm Ti1}_z , 0, m ^{\rm Mn}_y , m ^{\rm Mn}_z )^T$.
We expand on $\bm m ^{\rm Ti1}=-\bm m ^{\rm Mn}=\hat{\bm x}$, with $\hat{\bm x}\parallel[111]$, $\hat{\bm y}\parallel[1\bar{1}0]$, and $\hat{\bm z}=\hat{\bm x}\times\hat{\bm y}$, keeping each vector normalized. At fixed $\mu=0$, $k_{\rm B}T=10\;{\rm meV}$, and
$N_k=128^3$, we obtain $n_e=2.00135$ and
\begin{align}
    C =
    \begin{pmatrix}
    0 & 0        & 0        & 0& 0 & 0\\
    0 & 28.5498 & 0        & 0& 28.5435 & 0 \\
    0 & 0        & 28.5498 & 0& 0 & 28.5435 \\
    0 & 0        & 0        & 0 & 0 & 0\\
    0 & 28.5435 & 0        & 0& 28.5552 & 0 \\
    0 & 0        & 28.5435 & 0& 0 & 28.5552
\end{pmatrix}
,
\end{align}
in ${\rm meV}$ per cell, where
$C_{ij}=\partial_i\partial_j\Phi_{\rm el}|_{\delta \bm m = 0}$.
With $\delta\bm m ^a=(m_y^a,m_z^a)$,
\begin{align}
    F_0
    = &
    \mathcal{J} \bm m ^{\rm Ti1} \cdot \bm m ^{\rm Mn}
    - \frac{K_{\rm Ti1} }{2} (m_x ^{\rm Ti1}) ^2 - \frac{K_{\rm Mn} }{2} (m_x ^{\rm Mn}) ^2
    \notag \\
    \approx &
    - \mathcal{J} - \frac{K_{\rm Ti1} }{2} - \frac{K_{\rm Mn} }{2}
    \notag \\
    &
    + \frac{\mathcal{J} + K_{\rm Ti1} }{2} |\delta \bm m ^{\rm Ti1} |^2
    + \frac{\mathcal{J} + K_{\rm Mn} }{2} |\delta \bm m ^{\rm Mn} |^2
    \notag \\
    &
    + \mathcal{J} \delta \bm m ^{\rm Ti1} \cdot \delta \bm m ^{\rm Mn}.
\end{align}
Thus $\mathcal{J}=C_{25}=C_{36}$,
$K_{\rm Ti1}=C_{22}-\mathcal{J}$, and
$K_{\rm Mn}=C_{55}-\mathcal{J}$ give
\begin{align}
    \mathcal{J} &= 28.5435 \; {\rm meV}, \\
    K_{\rm Ti1} &= 6.2997 \; \mu{\rm eV}, \\
    K_{\rm Mn} &= 11.6624 \; \mu{\rm eV},
\end{align}
per cell; the two anisotropies are kept distinct.
For $\bm B_{\rm ext}=B_{\rm ac}(0,-\sin\Omega t,\cos\Omega t)$,
we solve the coupled LLG equations
\begin{align}
    \dot {\bm m } ^a &= - \gamma \bm m ^a \times \bm B ^a_{\rm eff} + \alpha \bm m ^a \times \dot {\bm m} ^a, \\
    \bm B ^a _{\rm eff} &= \bm B _{\rm ext} - \frac{1}{\mu^a } \frac{\partial F_0 }{\partial \bm m ^a}.
\end{align}

Setting $\bm n\equiv(\bm m ^{\rm Ti1}-\bm m ^{\rm Mn})/2$,
$\bm N\equiv(\bm m ^{\rm Ti1}+\bm m ^{\rm Mn})/2$, and linearizing about the antiparallel state, we use
\begin{align}
    u \equiv n_y + i n_z &= u_{\Omega} e^{i \Omega t }, \\
    v \equiv N_y + i N_z &= v_{\Omega} e^{i \Omega t }.
\end{align}

The numerical steady-state solution thus gives
\begin{align}
    n_\perp \equiv &|u _\Omega | = 2.63519 \times 10^{-3}, \\
    \Omega_{\rm res} = & 8.364 \times 10^{11} \; {\rm rad/s},
\end{align}
where $\Omega_{\rm res}$ maximizes $n_\perp$ for the stated rotation
direction. Note that $\Omega_{\rm res}/2\pi=133.12\;{\rm GHz}$.
We use $\gamma=2\mu_B/\hbar$ and $\mu^{\rm Ti1} = \mu ^{\rm Mn} = 2.59 \mu_B, \; \alpha = 0.01, \;B _{\rm ac} = 10 \; {\rm mT}$ \cite{Jakobsson2015}.

\subsection{Spin-Induced Current in Ti${}_2$MnAl}

In the sublattice basis, the exchange-coupling matrix is written as
\begin{align}
    \widehat J
    &=(J_0-J_3)\frac{\tau_0-\tau_3}{2}
    +(J_0+J_3)\frac{\tau_0+\tau_3}{2}
    \notag \\
    &\equiv
    J_{\rm exc}^{\rm Ti1}\frac{\tau_0-\tau_3}{2}
    +J_{\rm exc}^{\rm Mn}\frac{\tau_0+\tau_3}{2}.
\end{align}
With $P_{\rm Ti1}=(\tau_0-\tau_3)/2$ and
$P_{\rm Mn}=(\tau_0+\tau_3)/2$, the exchange Hamiltonian is
\begin{align}
    H_{\rm exc}
    =-\sum_{a={\rm Ti1},{\rm Mn}}
    J_{\rm exc}^a P_a\otimes(\bm m^a\cdot\bm\sigma).
\end{align}
The parameter values in Eq.~\eqref{eq:parameters} give
$J_0=1.1995\;{\rm eV}$ and $J_3=-0.4159\;{\rm eV}$.

For normalized spinors satisfying
$(\bm b_a\cdot\bm\sigma)\ket{\chi_{a,\pm}}
=\pm|\bm b_a|\ket{\chi_{a,\pm}}$, the two degenerate states are
\begin{align}
    \ket{u_1} =
    \begin{pmatrix}
        \ket{\chi_{{\rm Ti1}, -}}\\
        \vec{0}_2
    \end{pmatrix},\;
    \ket{u_2} =
    \begin{pmatrix}
        \vec{0}_2 \\
        \ket{\chi_{{\rm Mn}, +}}
    \end{pmatrix},
\end{align}
where $\vec{0}_2$ is a $2 \times 1$ zero vector. Projecting onto this
two-dimensional eigenspace gives
\begin{align}
    U^\dagger H_{\rm TMA} (\bm k_W; \hat{\bm x}, - \hat{\bm x}) U
    =
    \begin{pmatrix}
        E_W  & 0 \\
        0 & E_W 
    \end{pmatrix},
\end{align}
where the $4\times2$ matrix
\begin{align}
    U =
    \begin{pmatrix}
        \ket{\chi_{{\rm Ti1}, -}} & \vec{0}_2 \\
        \vec{0}_2 & \ket{\chi_{{\rm Mn}, +}}
    \end{pmatrix}
\end{align}
satisfies $U^\dagger U=I_2$, and the Weyl-node energy is
\begin{align}
    E_W = \xi_{\rm Ti1}-|\bm b_{\rm Ti1}|
    = \xi_{\rm Mn}+|\bm b_{\rm Mn}|.
\end{align}

To extract the N\'eel coupling, take $\delta\bm N=0$.
With $\bm q=\bm k-\bm k_W$ and the fixed equilibrium basis $U$,
the time-dependent model to first order in $\bm q$ and
$\delta\bm n =(0, \delta n_y,\delta n_z)^T$ is
\begin{align}
    H_{\bm k \cdot \bm p}
    &\equiv U^\dagger
    H_{\rm TMA} (\bm k_W+\bm q; \bm m^{\rm Ti1}, \bm m^{\rm Mn}) U
    \notag \\
    &\approx E_W I_2
    +\sum_i q_i U^\dagger (\partial_{k_i}H_0) U
    +\sum_{\beta=y,z}\delta n_\beta(t)\Gamma_\beta,
\end{align}
where $H_0(\bm k)=H_{\rm TMA}(\bm k;\hat{\bm x},-\hat{\bm x})$
and all derivatives are evaluated at $\bm k_W$. The exchange vertices are
\begin{align}
    \Gamma_\beta & \equiv U^\dagger
    \begin{pmatrix}
        -J_{\rm exc}^{\rm Ti1}\sigma_\beta & 0 \\
        0 & +J_{\rm exc}^{\rm Mn}\sigma_\beta
    \end{pmatrix} U
    \notag \\
    &= 
    \begin{pmatrix}
        \operatorname{Tr} [\Gamma_\beta/2] + J_{{\rm exc}, \beta}^{\rm eff} & 0 \\
        0 & \operatorname{Tr} [\Gamma_\beta/2] - J_{{\rm exc}, \beta}^{\rm eff} 
    \end{pmatrix}
    , \\
    J_{{\rm exc}, \beta}^{\rm eff}
    &\equiv \frac{J_{\rm exc}^{\rm Ti1}\hat b_{{\rm Ti1},\beta}
    -J_{\rm exc}^{\rm Mn}\hat b_{{\rm Mn},\beta}}{2},
\end{align}
where $\hat{\bm b}_a=\bm b_a/|\bm b_a|$.

For $\hat{\bm x}\parallel[111]$, $\hat{\bm y}\parallel[1\bar10]$,
$\hat{\bm z}=\hat{\bm x}\times\hat{\bm y}$, we choose
\begin{align}
    \frac{\bm k_W}{2\pi}=(-0.7641737293,-0.5188957639,0)
\end{align}
in cubic crystal coordinates with lattice constant set to unity.
We obtain $E_W=-1.89413\;{\rm meV}$ and
\begin{align}
    \begin{pmatrix}
        J_{{\rm exc}, x}^{\rm eff} \\
        J_{{\rm exc}, y}^{\rm eff} \\
        J_{{\rm exc}, z}^{\rm eff}
    \end{pmatrix}
    =
    \begin{pmatrix}
        0  \\
        0.0716059  \\
        -0.0172105
    \end{pmatrix}\;{\rm eV}.
\end{align}
Therefore, defining the effective exchange coupling as the maximum
coupling to a unit transverse displacement,
\begin{align}
    J_{\rm exc}^{\rm eff} &\equiv 
    \sqrt{
    (J_{{\rm exc}, x}^{\rm eff})^2 +
    (J_{{\rm exc}, y}^{\rm eff})^2+(J_{{\rm exc}, z}^{\rm eff})^2} 
    \notag \\
    &= 0.0736452\;{\rm eV}.
\end{align}
For circular N\'eel precession of amplitude $n_\perp$, the length of the axial gauge field is $J_{\rm exc}^{\rm eff}n_\perp/(\sqrt{2} v)$.

\subsection{Estimation of the Spin-Induced Current}

\begin{figure}[tb]
    \centering
    \includegraphics[width=1\linewidth]{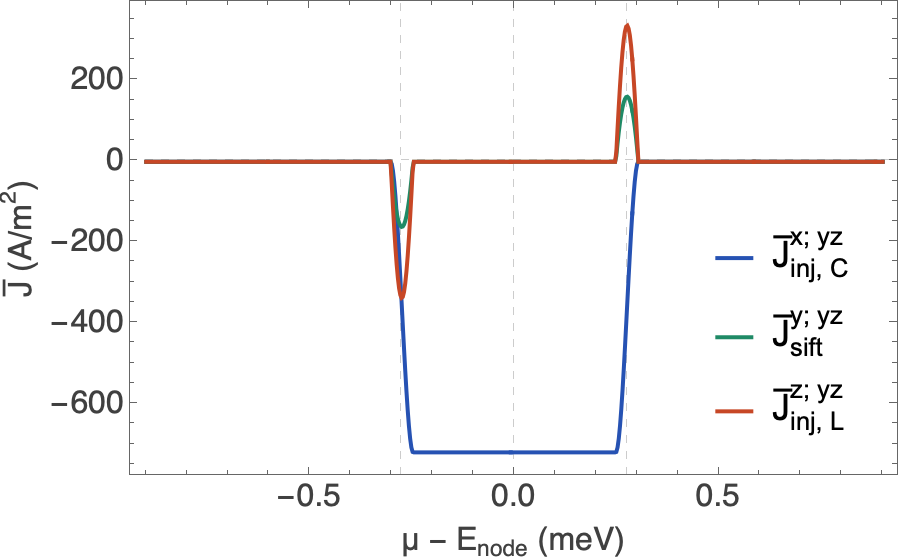}
    \caption{Chemical-potential dependence of the spin-induced dc current
    density in the single-cone isotropic-exchange approximation for
    circular precession in the $yz$ plane and a tilt along $z$.
    Blue, green, and red curves denote the circular-injection current
    $\bar J_{{\rm inj},C}^{x;yz}$, shift current
    $\bar J_{\rm shift}^{y;yz}$, and linear-injection current
    $\bar J_{{\rm inj},L}^{z;yz}$, respectively.
    The chemical potential is measured from the Weyl-node energy
    $E_{\rm node}$; the parameters are specified in the text.}
    \label{fig:ti2mnal_estimation}
\end{figure}

\begin{table*}[htbp]
\caption{
Correspondence between the electromagnetic gauge field and the chiral
gauge field.
The particle-current convention with $e=\hbar=1$ is used.
Here,
\(
\int_{\bm k}
\equiv
\int\frac{d\bm k}{(2\pi)^3}
\)
and
\(
\mathcal Q_{12}^{xy}
\equiv
\mathcal G_{12}^{xx}
+
\mathcal G_{12}^{yy}
-
\mathcal F_{12}^{xy}.
\)
The optical formulas have been rewritten in the particle-current convention and in the relaxation-time convention used in this work.
}
\label{tab:gauge_chiral}

\renewcommand{\arraystretch}{1.8}
\setlength{\tabcolsep}{4pt}

\begin{tabular}{ccc}
\toprule
&
Electromagnetic gauge field \cite{ahn,juan2017,wy2021}
&
Chiral gauge field (our work)
\\
&
\parbox[c]{0.41\textwidth}{
\centering
\(
\displaystyle
\begin{aligned}
	\bm A_{\Omega}
	&=
	A (1,i,0)^T,
	\;
	\bm E_{\Omega}
	=
	i\Omega\bm A_{\Omega},
	\; 
	E
	\equiv
	|\bm E_{\Omega}|
	=
	\sqrt{2} \Omega A
\end{aligned}
\)
}
&
\parbox[c]{0.41\textwidth}{
\centering
\(
\displaystyle
\begin{aligned}
	\bm A_{5,\Omega}
	&=
	\frac{A_5}{\sqrt{2}}
	(1,i,0)^T,
	\;
	A_5
	=
	\frac{J_{\rm exc}m_\perp}{\sqrt{2}v}, \;
    E_5 \equiv \Omega A_5
\end{aligned}
\)
}
\\

\hline
\hline

\parbox[c]{0.11\textwidth}{
\centering
Injection
}
&
\parbox[c]{0.41\textwidth}{
\centering
\(
\displaystyle
\begin{aligned}
	\bar J_{\rm inj}^{a;xy}(E) \!
	&= \!
	4 \pi\tau E^2\!\! \!
	\int_{\bm k}\!\! \!
	f_{12}
	(v_{11}^a \!- \!v_{22}^a)
	\mathcal Q_{12}^{xy}
	\delta (\epsilon_{12} \! + \!\Omega)
\end{aligned}
\)
}
&
\parbox[c]{0.41\textwidth}{
\centering
\(
\displaystyle
\begin{aligned}
    \bar J_{\rm inj}^{a;xy}(A_5)
	& = \!
	2\pi\tau E_5 ^2 \!\!\!
	\int_{\bm k} \!\! \!
	f_{12}
	(v_{11}^a \! - \!v_{22}^a )
	\mathcal Q_{12}^{xy}
	\delta ( \epsilon_{12} \!+ \!\Omega)
\end{aligned}
\)
}
\\

\parbox[c]{0.11\textwidth}{
\centering
Shift
}
&
\parbox[c]{0.41\textwidth}{
\centering
\(
\displaystyle
\begin{aligned}
	\bar J_{\rm shift}^{a;xy}(E)
	&=
	2 {\pi E^2}
	\int_{\bm k}
	f_{12}
	R_{12, -}^{a;xy}
	\mathcal Q_{12}^{xy}
	\delta (\epsilon_{12}+\Omega)
\end{aligned}
\)
}
&
\parbox[c]{0.41\textwidth}{
\centering
\(
\displaystyle
\begin{aligned}
	\bar J_{\rm shift}^{a;xy}(A_5)
	&=
	\pi E_5 ^2
	\int_{\bm k}
	f_{12}
	R_{12, -}^{a;xy}
	\mathcal Q_{12}^{xy}
	\delta ( \epsilon_{12}+\Omega )
\end{aligned}
\)
}
\\
\bottomrule
\end{tabular}

\end{table*}

We estimate a reference current scale using a single tilted Weyl cone
with an isotropic exchange approximation. We use the magnitude
$J_{\rm exc}^{\rm eff}$ obtained above and identify $m_\perp=n_\perp$.
The parameters are
\begin{align}
    v &= 6.0 \times 10^5 \; {\rm m/s}, \quad
    J_{\rm exc}^{\rm eff} = 73.65 \; {\rm meV}, \notag  \\
    n_\perp &= 2.63519 \times 10^{-3}, \quad
    \tau = 1.0 \; {\rm ps}, \quad \eta = +1, \\
    \hbar\Omega &= 0.550528 \; {\rm meV}, \quad
    \Gamma = \frac{\hbar}{2\tau} = 0.329106 \; {\rm meV}.
    \notag 
\end{align}
The corresponding chiral gauge-field amplitude is
\begin{align}
    A_5 = \frac{J_{\rm exc}^{\rm eff} n_\perp}{\sqrt{2}\hbar v}
    = 3.47498 \times 10^5 \; {\rm m^{-1}}.
\end{align}

Fig.~\ref{fig:ti2mnal_estimation} shows the chemical-potential
dependence relative to the Weyl-node energy $E_{\rm node}$.
The circular-injection current $\bar J_{{\rm inj},C}^{x;yz}$
reaches a negative plateau of approximately
$-7.2\times10^2\;{\rm A/m^2}$.
The linear-injection and shift currents,
$\bar J_{{\rm inj},L}^{z;yz}$ and $\bar J_{\rm shift}^{y;yz}$,
have peak values of approximately
$\pm3.4\times10^2\;{\rm A/m^2}$ and
$\pm1.6\times10^2\;{\rm A/m^2}$, respectively, near
$\mu-E_{\rm node}=\pm\hbar\Omega/2=\pm0.275264\;{\rm meV}$.
Both are positive on the electron side and negative on the hole side.

These are formal second-order model estimates: the scalar approximation does not retain the rank-one projected coupling, and
$J_{\rm exc}^{\rm eff}n_\perp/\Gamma\approx 0.590$ and
$\Gamma/(\hbar\Omega)\approx 0.598$ do not establish a well-separated perturbative regime.

\section{Discussion}

Let us discuss comparing our work with photoinduced currents.
In noncentrosymmetric systems, nonlinear current responses induced by optical electric fields have been extensively investigated. The theoretical framework for these responses is now well established, and their topological and geometrical aspects have been clarified in terms of Berry curvature, Berry connection, and quantum metric  \cite{ss2000, mn2016, sf2015, juan2017, ck}.  
In Weyl semimetals with broken $\cal P$ symmetry, which host topological Weyl nodes and exhibit large Berry curvature around these nodes, nonlinear responses have been well investigated and classified from the perspective of quantum geometry \cite{wy2021, ahn}.
While these geometric classifications are highly successful, they primarily consider conventional electromagnetic fields as the driving force.
This phenomenon is prominent in MWSMs due to the strong exchange coupling between localized spins and conduction electrons. 
We find that the spin-induced response consists of a shift current proportional to $\Omega^1$ and an injection current proportional to $\Omega^2\tau$, drawing a clear contrast to the low-frequency behavior of conventional photogalvanic effects \cite{ahn, juan2017} (see Table \ref{tab:gauge_chiral}).

\section{Summary}


In this study, we evaluated the nonlinear DC current induced by steady spin precession in a simple model of a MWSM using the Floquet–Keldysh Green’s function method. We showed that both circular shift and circular injection currents contribute to the response coefficient.
In the untilted case, the total current from the two Weyl nodes exhibits discontinuous changes as a function of the driving frequency $\Omega$. This behavior reflects three distinct regimes: (i) transitions at both nodes are forbidden because the final states lie below the chemical potential, resulting in zero current; (ii) transitions are allowed at both nodes, but their current contributions cancel exactly, also resulting in zero net current; and (iii) transitions are allowed at only one node, yielding a finite net current.
Introducing a tilt to the Weyl cones causes the resonant surface to incline, so that transitions become allowed progressively as the driving frequency changes. Consequently, the discontinuous jumps observed in the untilted model are substantially smoothed out. Furthermore, when transitions are allowed over only part of the resonant surface because some initial states are unoccupied or some final states are occupied, the reduced symmetry of the tilted Weyl cones becomes particularly evident. In this regime, additional nonlinear current responses, including shift currents and linear injection currents, are allowed.

\section{ACKNOWLEDGMENTS}

This work was supported by JSPS KAKENHI Grant Numbers JP24H00197, JP25H01250, JP26K00652, and JP26K00660.

\bibliography{ref}

\appendix

\section{Floquet-Keldysh Formalism}
\label{FK}

To describe the nonequilibrium steady state, we employ the Floquet-Keldysh Green's function formalism.\cite{mn2016, kamenev, Matsyshyn2021} 
Here, the operators $a^\dagger$ ($a$) and $b^\dagger$ ($b$) denote the creation (annihilation) operators for the system and the bath, respectively.
As illustrated in Fig.~\ref{fig:fig1}(a), the total Hamiltonian of the system is given by
\begin{align}
    {\cal H}_{\rm tot} (t) &= {\cal H}(t) + {\cal H}_{\rm bath} + {\cal H}_{\rm tun}, \\
    {\cal H}(t) &= \sum_{\alpha\neq \beta} 
    a^\dagger_\alpha {\cal H}^s_{\alpha \beta} (t) a_\beta ={\cal H}_0 + V(t), \\ 
    {\cal H}_{\rm bath} &= \sum_{\alpha} {\cal E}^{\rm bath}_\alpha b^\dagger_\alpha b_\alpha, \\ 
    {\cal H}_{\rm tun} &= \sum_{\alpha, \beta} ({\cal T}^\alpha b^\dagger_\alpha a_\beta + {\rm h.c.}),
\end{align}
where the indices $\alpha$ and $\beta$ denote the relevant band indices.

To formulate the problem using path integrals, we define the partition function and the corresponding action as
\begin{align}
    {\cal Z} &= \int {\cal D}[a,a^\dagger,b,b^\dagger] e^{iS_{\rm tot}}, \\
    S_{\rm tot} &= \int_C dt ({\cal L} + {\cal L}_{\rm bath} + {\cal L}_{\rm tun}), \\
    {\cal L} &= \sum_{\alpha, \beta} 
    a^\dagger_\alpha( i \partial _t - {\cal H}_{\alpha \beta} (t))a_\beta,  \\
    {\cal L}_{\rm bath} & = \sum_{\alpha} b^\dagger_\alpha (i\partial _t - {\cal E}^{\rm bath}_\alpha) b_\alpha  , \\
    {\cal L}_{\rm tun} & = - \sum_{\alpha, \beta} ({\cal T}_\alpha b^\dagger_\alpha  a_\beta + {\rm h.c.}) ,  \\
    \braket{b^\dagger _\alpha b _\alpha} &= f({\cal E}^{\rm bath}_\alpha).
\end{align}
Here, $\mathcal E_\alpha^{\rm bath}$ denotes the energy of bath state $\alpha$. In the wide-band approximation, the bath-induced broadening is characterized by the relaxation rate $\Gamma$ defined below.

\subsection{Green's function}

To systematically calculate the physical observables and correlation functions, we introduce the contour-ordered Green's functions via the generating functional approach.

We define the generating functional $W[\bar J, J, \bar K, K]$ by introducing auxiliary Grassmann source fields $\bar J, J$ that couple to the system operators $a, a^\dagger$, and $\bar K, K$ that couple to the bath operators $b, b^\dagger$:
\begin{align}
    &e^{i W[\bar J, J, \bar K, K]} = 
    \notag \\
    &\int \mathcal{D}[a,a^\dagger] \mathcal{D}[b,b^\dagger] \notag \\
    \times &\ \exp \Bigg[iS - i \int_C dt \sum_\alpha \left( \bar J_\alpha a_\alpha + a^\dagger_\alpha J_\alpha + \bar K_\alpha b_\alpha + b^\dagger_\alpha K_\alpha \right) \Bigg].
\end{align}

By taking the second functional derivatives of the generating functional with respect to these source fields and subsequently setting them to zero, we obtain the full interacting contour-ordered Green's functions for the system and the bath:
\begin{align}
    G^{\mu \nu} (t,t') 
    &= -i \frac{\delta^2 e^{i W[\bar J, J, \bar K, K]}}{\delta \bar J_\mu (t) \delta J_\nu (t')} \Bigg|_{\bar J, J, \bar K, K = 0}
    \notag \\
    &= -i \Braket{T_C a_\mu(t) a_\nu^\dagger(t')} ,
    \\
    G^{\mu \nu}_{\rm bath} (t,t') 
    &= -i \frac{\delta^2 e^{i W[\bar J, J, \bar K, K]}}{\delta \bar K_\mu (t) \delta K_\nu (t')} \Bigg|_{\bar J, J, \bar K, K = 0}
    \notag \\
    &= -i \Braket{T_C b_\mu(t) b_\nu^\dagger(t')} .
\end{align}
Here, $T_C$ denotes the contour-ordering operator, which orders the field operators along the Keldysh contour $C$. The expectation value $\braket{(\cdots)}$ with respect to the full action $S$ is defined as
\begin{equation}
    \braket{(\cdots)} = \frac{1}{\cal Z} \int \mathcal{D}[a,a^\dagger] \mathcal{D}[b,b^\dagger] e^{iS_{\rm tot}} (\cdots).
\end{equation}
Similarly, we define the unperturbed (bare) Green's functions by taking the expectation value with respect to the non-interacting action $S_0$ (i.e., the action decoupled from the tunneling Hamiltonian ${\cal H}_{\rm tun}$):
\begin{align}
    g^{\mu \nu} (t,t') 
    &= -i \Braket{T_C a_\mu(t) a_\nu^\dagger(t')}_0 ,
    \\
    g^{\mu \nu}_{\rm bath} (t,t') 
    &= -i \Braket{T_C b_\mu(t) b_\nu^\dagger(t')}_0 ,
\end{align}
where the subscript $0$ indicates that the path integral is performed using $S_0 = \int_C dt \sum_{\alpha, \beta} a^\dagger_\alpha( i \partial _t - {\cal H}_{0, \alpha \beta} )a_\beta$:
\begin{equation}
    \braket{(\cdots)} = \frac{1}{\cal Z} \int \mathcal{D}[a,a^\dagger] \mathcal{D}[b,b^\dagger] e^{iS_{0}} (\cdots).
\end{equation}

Depending on the relative positions of the time arguments $t$ and $t'$ on the forward or backward branches of the closed Keldysh contour, these contour-ordered Green's functions can be analytically continued into real-time components. This mapping naturally yields the lesser ($G^<$), greater ($G^>$), retarded ($G^R$), and advanced ($G^A$) Green's functions, which are essential for describing the non-equilibrium dynamics and spectral properties of the driven system.

\subsection{Time Periodicity of the Action}

Since the system Hamiltonian is time-periodic with period $T = 2\pi/\Omega$ such that ${\cal H} (t) = {\cal H}(t + T)$ due to the external driving $V(t) = V(t+T)$, we can expand it as a discrete Fourier series:
\begin{align}
    {\cal H}_{\alpha \beta} (t) 
    &= \sum_{m = -\infty}^\infty H_m^{\alpha \beta} e^{-im\Omega t}
    , \\ 
    H_m^{\alpha \beta} 
    &= \int_0^T \frac{dt}{T} e^{+im \Omega t} {\cal H}^{\alpha \beta} (t),
\end{align}
where the field operators are expressed in the Schr\"odinger picture. By introducing the continuous Fourier transforms for these operators,
\begin{equation}
    a^\dagger_\alpha(t) = \int \frac{d\omega}{2\pi} e^{i\omega t} a^\dagger_\alpha(\omega), \quad
    a_\beta(t) = \int \frac{d\omega'}{2\pi} e^{-i\omega' t} a_\beta(\omega'),
\end{equation}
the action can be expanded and transformed into the frequency-space representation. Noting that the time integration is performed over the Keldysh contour $C$, the field operators implicitly carry the Keldysh branch indices ($+$ and $-$). To express the action in the Floquet basis, we fold the frequency integral into the Floquet Brillouin zone $[0, \Omega)$ using the relation:
\begin{equation}
    \int_{-\infty}^\infty d\omega \chi (\omega) = \sum_{n = -\infty}^\infty
    \int_{0}^\Omega d\omega \chi(\omega + n\Omega),
\end{equation}
where $\chi(\omega)$ is an arbitrary function. By applying this identity and introducing the Floquet component notation $a_{\alpha, n}^{(\dagger)}(\omega) \equiv a^{(\dagger)}_\alpha(\omega + n\Omega)$, the action is rewritten as
\begin{align}
    S &=
    \int_C dt \sum_{\alpha, \beta} 
    a^\dagger_\alpha( i \partial _t - \sum_{m} H_m^{\alpha \beta} e^{-im\Omega t} )a_\beta
    \notag \\
    &= \sum_{\alpha, \beta}\sum_{l,n} \int_{0}^\Omega d\omega \, a^\dagger_{\alpha, n}(\omega)
    \Big( \omega \delta_{l,n} - (H_F^{\alpha \beta})_{n,l} \Big) a_{\beta, l}(\omega),
\end{align}
where we have shifted the summation index to $l = m+n$ and defined the Floquet Hamiltonian matrix elements as 
\begin{equation}
    (H_F^{\alpha \beta})_{n,l} \equiv H_{l-n}^{\alpha \beta} - n\Omega\delta_{l,n}.
\end{equation}
Similarly, the actions for the tunneling coupling and the bath can be transformed into the Floquet basis:
\begin{align}
    S_{\rm tnn}
    & =
    \sum_{\alpha, \beta} \sum_n \int_{0}^\Omega d\omega
    {\cal T}_\alpha
    b^\dagger_{\alpha, n} (\omega)
    a_{\beta, n} (\omega)
    + {\rm h.c.}, 
    \\
    S_{\rm bath} & =
    \sum_{\alpha} \sum_n \int_0^\Omega d \omega
    b^\dagger_{\alpha, n} (\omega) 
    (\omega+ n\Omega - {\cal E}^{\rm bath}_\alpha) b_{\alpha,n } (\omega)
    \notag \\
    & \equiv 
    \sum_{\alpha} \sum_n \int_0^\Omega d \omega
    b^\dagger_{\alpha, n} (\omega) 
    \Big(g_{{\rm bath}} ^{-1} \Big)^\alpha_n (\omega) b_{\alpha,n } (\omega),
\end{align}
where we define the inverse bare Green's function for the bath as
\begin{align}
    \Big(g_{{\rm bath}} ^{-1} \Big)^\alpha_n (\omega)
    =
    \omega+ n\Omega - {\cal E}^{\rm bath}_\alpha.
\end{align}
Here, $g_{{\rm bath}} ^{-1}$ acts as the inverse operator of the bare bath Green's function $g_{{\rm bath}}$, satisfying the relation:
\begin{align}
    \Big({g_{{\rm bath}} ^{-1}} \Big)^\alpha_n (\omega) g_{{\rm bath}, m}^\beta (\omega')
    = \delta_{\alpha \beta} \delta_{nm} \delta(\omega - \omega').
\end{align}
The other inverse Green's functions, $G^{-1}, G^{-1}_{\rm bath}$, and $g^{-1}$, are defined in a similar manner.

\subsection{Effective Action}
By completing the square and integrating out the bath fields $b^\dagger$ and $b$, we obtain the effective interaction action:
\begin{align}
    &S_{\rm int} \equiv 
    \notag \\
    &- \sum_n \sum_{\alpha, \beta, \gamma} \int_{0}^\Omega d\omega
    a_{\gamma, n}^\dagger (\omega) 
    |{\cal T}_{\alpha}|^2 g_{{\rm bath}, n}^\alpha (\omega)
    a_{\beta, n}(\omega).
\end{align}
Therefore, the total effective action of the system becomes
\begin{align}
    &S + S_{\rm int} = 
    \notag \\
    &\sum_n \sum_{\alpha, \beta, \gamma} \int_{0}^\Omega d\omega
    a_{\gamma, n}^\dagger
    (\delta_{\beta, \gamma} g_n^{-1} (\omega) 
    - |{\cal T}_\alpha|^2 g_{{\rm bath}, n}^\alpha (\omega) )
    a_{\beta, n}.
\end{align}

From this expression, we identify the self-energy induced by the bath as
\begin{align}
    \Sigma_n(\omega)
    &= \sum_\alpha |\mathcal{T}_\alpha|^2 g_{\mathrm{bath}, n}^\alpha (\omega)
    \notag \\
    &= \sum_\alpha\frac{|\mathcal{T}_\alpha|^2 }{\omega + n\Omega - \mathcal{E}^{\mathrm{bath}}_\alpha}.
\end{align}
To evaluate the retarded component of the self-energy, which guarantees the causality of the system's response, we must perform an analytic continuation to the upper half of the complex frequency plane. Practically, this is achieved by introducing an infinitesimally small positive imaginary part $i\eta$ ($\eta \to 0^+$) into the denominator of the bare bath Green's function, yielding
\begin{align}
    \Sigma^R_n(\omega) 
    &= \sum_\alpha |\mathcal{T}_\alpha|^2 g_{\mathrm{bath}, n}^{R, \alpha}(\omega) \notag \\
    &= \sum_\alpha \frac{|\mathcal{T}_\alpha|^2}{\omega + n\Omega - \mathcal{E}^{\mathrm{bath}}_\alpha + i\eta}.
\end{align}
Using the identity ${1}/{(x + i\eta)} = \mathcal{P}\left({1}/{x}\right) - i\pi \delta(x)$, where $\mathcal{P}$ denotes the Cauchy principal value, the self-energy can be separated into its real and imaginary parts:
\begin{align}
    &\Sigma^R_n(\omega) = 
    \notag \\
    & \mathcal{P} \sum_\alpha \frac{|\mathcal{T}_\alpha|^2}{\omega + n\Omega - \mathcal{E}^{\mathrm{bath}}_\alpha} 
    - i\pi \sum_\alpha |\mathcal{T}_\alpha|^2 \delta(\omega + n\Omega - \mathcal{E}^{\mathrm{bath}}_\alpha).
\end{align}

Here, we introduce the wide-band approximation. This approximation assumes that the density of states of the bath and the coupling strength $|\mathcal{T}_\alpha|^2$ are constant with respect to energy, and that the bandwidth is infinitely wide. 
Under this approximation, the real part given by the principal value integral can be neglected, as it simply renormalizes the system Hamiltonian by providing a constant shift to the energy. 
Furthermore, the imaginary part can be regarded as constant with respect to variations in the Floquet energy $n\Omega$ and $\omega$. Thus, we can define an energy-independent relaxation rate $\Gamma$ as follows:
\begin{equation}
    \Gamma \equiv 2\pi \sum_\alpha |\mathcal{T}_\alpha|^2 \delta(\omega - \mathcal{E}^{\mathrm{bath}}_\alpha).
\end{equation}
Consequently, in the wide-band limit, the retarded self-energy of the system reduces to an extremely simple form as a purely imaginary constant:
\begin{equation}
    \Sigma^R_n(\omega) \approx -i \frac{\Gamma}{2}.
\end{equation}

With the retarded self-energy determined, the general lesser Green's function can be evaluated. The $z$-component of the current density is expressed in terms of this lesser Green's function as
\begin{equation}
    J(t)
    =
    - i
    \int \frac{d\bm k}{(2\pi)^3}
    \int \frac{d\omega}{2\pi}
    {\rm Tr}
    \left[
    \tilde{v}^z(\bm{k})
    \bm{G}^<(\bm{k},\omega)
    \right],
\end{equation}
where the trace ${\rm Tr}$ is taken over both the orbital/spin and Floquet indices, and $\bm{G}^<(\omega)$ is determined by the Dyson equation
\begin{equation}
    \bm{G}^<(\omega) = \bm{G}^R(\omega)\bm{\Sigma}^<(\omega)\bm{G}^A(\omega).
    \label{eq:dyson_general}
\end{equation}

\begin{widetext}
\section{Derivation of the Geometric Representation of the Nonlinear Current}
\label{app:geometric_current}

In this Appendix, we rewrite the contributions to the DC current
in terms of the geometric quantities of the Bloch bands.
Throughout this paper, we use the particle-current convention and omit
the overall electric-charge factor.

We first consider the shift current.
Using the relation between the interband velocity matrix element and the
non-Abelian Berry connection,
$v_{12}^{\alpha}=-i\epsilon_{12}\mathcal A_{12}^{\alpha}$,
we evaluate the momentum derivative of the interband velocity matrix
element as
\begin{align}
	\partial_a v_{12}^{b}
	&=
	\partial_a \bra{u_1}\hat v^{b}\ket{u_2}
	\notag \\
	&=
	\bra{\partial_a u_1}\hat v^{b}\ket{u_2}
	+
	\bra{u_1}\partial_a \hat v^{b}\ket{u_2}
	+
	\bra{u_1}\hat v^{b}\ket{\partial_z u_2}
	\notag \\
	&=
	-i\sum_{l=1,2}\mathcal A_{1l}^{a}v_{l2}^{b}
	+
	\Big(\partial_a \hat v^{b}\Big)_{12}
	+
	i\sum_{l=1,2}v_{1l}^{b}\mathcal A_{l2}^{a}
	\notag \\
	&=
	\Big(\partial_a \hat v^{b}\Big)_{12}
	-i (\mathcal A_{11}^{a}-\mathcal A_{22}^{a} )v_{12}^{b}
	+i (v_{11}^{b}-v_{22}^{b} )\mathcal A_{12}^{a}.
	\label{eq:appB_velocity_derivative}
\end{align}
Rearranging Eq.~\eqref{eq:appB_velocity_derivative}, we obtain
\begin{align}
	\Big(\partial_a \hat v^{b}\Big)_{12}
	&=
	\Big[
		\partial_a 
		+i (\mathcal A_{11}^{a}-\mathcal A_{22}^{a } )
	\Big]v_{12}^{b}
	+i\mathcal A_{12}^{a }
	(v_{22}^{b}-v_{11}^{b}).
	\label{eq:appB_covariant_velocity_derivative}
\end{align}
Substituting Eq.~\eqref{eq:appB_covariant_velocity_derivative} into the
numerator of $\bar J_{\rm shift}^{a ;bc }$ gives
\begin{align}
	\tilde v_b ^a h_c -\tilde v_c ^a h_b
	&=
	{\rm Im}\Big[
		(\tilde v_b ^a -i\tilde v_c ^a )
		(h_b +ih_c )
	\Big]
	\notag \\
	&=
	\frac{A_5^2}{2}
	{\rm Im}\Big[
		\Big(
			(\partial_a \hat v^b )_{12}
			-i (\partial_a \hat v^c )_{12}
		\Big)
		(v_{21}^b +iv_{21}^c )
	\Big]
	\notag \\
	&=
	\frac{A_5^2}{2}
	{\rm Im}\Big[
		\Big(
			\{
				\partial_z
				+i(\mathcal A_{11}^a -\mathcal A_{22}^a )
			\}
			(v_{12}^b -iv_{12}^c )
			-i\mathcal A_{12}^a 
			\{
				(v_{11}^b -v_{22}^b )
				-i (v_{11}^c -v_{22}^c )
			\}
		\Big)
		(v_{12}^b -iv_{12}^c )^*
	\Big]
	\notag \\
	&=
	\frac{A_5^2}{2}
	\left|v_{12}^b -iv_{12}^c \right|^2
	\Big[
		\partial_a \arg(v_{12}^b -iv_{12}^c )
		+\mathcal A_{11}^a -\mathcal A_{22}^a 
	\Big]
	-
	\frac{A_5^2}{2}
	{\rm Re}\Big[
		\mathcal A_{12}^a 
		\{
			(v_{11}^b -v_{22}^b )
			-i(v_{11}^c -v_{22}^c)
		\}
		(v_{12}^b -iv_{12}^c )^*
	\Big]
	\notag \\
	&=
	\frac{A_5^2}{2}
	\left|v_{12}^b -iv_{12}^c \right|^2
	R_{12}^{a ;bc }
	-
	\frac{A_5^2}{2}
	{\rm Re}\Big[
		\mathcal A_{12}^a 
		\{
			(v_{11}^b -v_{22}^b )
			-i(v_{11}^c -v_{22}^c )
		\}
		(v_{12}^b -iv_{12}^c )^*
	\Big],
	\label{eq:appB_shift_numerator}
\end{align}
where the shift vector is defined by
\begin{align}
	R_{12,\pm}^{a ;bc }
	&\equiv
	\partial_a \arg(v_{12}^b \pm iv_{12}^c )
	+\mathcal A_{11}^a -\mathcal A_{22}^a .
\end{align}
For the Weyl Hamiltonian considered here, the quantity inside
${\rm Re}[\cdots]$ in the second term of
Eq.~\eqref{eq:appB_shift_numerator} is purely imaginary.
the second term therefore vanishes identically, and the shift-current contribution becomes
\begin{align}
	\bar J_{\rm shift}^{a ;bc }
	&=
	\pi A_5^2
	\int\frac{d\bm k}{(2\pi)^3}
	\left|v_{12}^b -iv_{12}^c \right|^2
	R_{12, -}^{a ;bc}
	f_{12}
	\delta\Big(\epsilon_{12}+\Omega\Big).
	\label{eq:appB_shift_geometric}
\end{align}

To evaluate the resonant momentum integral, we introduce cylindrical
coordinates,
$k_\perp\equiv\sqrt{k_x^2+k_y^2}$,
and define the resonant momentum $\bar k\equiv\Omega/(2v)$.
The delta function can then be rewritten as
\begin{align}
	\delta\Big(\epsilon_{12}+\Omega\Big)
	&=
	\delta\Big(\Omega-2vk\Big)
	\notag \\
	&=
	\delta\Big(
		\Omega-2v\sqrt{k_\perp^2+k_z^2}
	\Big)
	\Theta\Big(\bar k^2-k_z^2\Big)
	\notag \\
	&=
	\frac{\bar k}{2v\sqrt{\bar k^2-k_z^2}}
	\delta\Big(
		k_\perp-\sqrt{\bar k^2-k_z^2}
	\Big)
	\Theta\Big(\bar k^2-k_z^2\Big)
	\notag \\
	&=
	\frac{\bar k}{2vk_\perp}
	\delta\Big(
		k_\perp-\sqrt{\bar k^2-k_z^2}
	\Big)
	\Theta\Big(\bar k^2-k_z^2\Big),
	\label{eq:appB_delta_cylindrical}
\end{align}
where we have used
\begin{align}
	\delta\Big(g(x)\Big)
	&=
	\sum_i
	\frac{\delta(x-x_i)}{\left|g'(x_i)\right|},
	\label{eq:appB_delta_identity}
\end{align}
with $x_i$ denoting the simple roots satisfying $g(x_i)=0$.
In the present case, the relevant root in the domain $k_\perp\geq0$ is
\begin{align}
	k_\perp
	&=
	\sqrt{\bar k^2-k_z^2},
	\label{eq:appB_resonant_root}
\end{align}
which exists only for $|k_z|\leq\bar k$.

For spin precession in the $xy$ plane, Eq.~\eqref{eq:appB_delta_cylindrical} gives the shift current along the $z$ direction as
\begin{align}
	\bar J_{\rm shift}^{z;xy}
	&=
	\frac{A_5^2v^2}{8\pi^2}
	\int_{-\infty}^{\infty}dk_z
	\int_0^{\infty}k_\perp dk_\perp
	\int_0^{2\pi}d\phi
	\Big(1+s\frac{k_z}{k}\Big)^2
	R_{12, -}^{z;xy}
	f_{12}
	\delta\Big(\Omega-2vk\Big)
	\notag \\
	&=
	\frac{A_5^2\Omega}{16 \pi^2}
	\int_{-\bar k}^{\bar k}dk_z
	\int_0^{2\pi}d\phi
	\Big(1+s\frac{k_z}{\bar k}\Big)^2
	R_{12, -}^{z;xy}
	f_{12}
	\notag \\
	&=
	\frac{A_5^2\Omega}{16 \pi^2}
	\int_{-\bar k}^{\bar k}dk_z
	\int_0^{2\pi}d\phi
	\Big(1+s\frac{k_z}{\bar k}\Big)^2
	\Big[
		\partial_z(-\phi)
		+\mathcal A_{11}^z-\mathcal A_{22}^z
	\Big]
	f_{12}
	\notag \\
	&=0.
	\label{eq:appB_shift_xy_zero}
\end{align}
Thus, the $z$-directed shift current vanishes for spin precession in the
$xy$ plane.

Next, we consider the injection current.
The hybridization amplitude can be expressed in terms of the quantum
metric and Berry curvature.
In the perturbativ limit, the dispersive factor associated with the
real part of the Green's function approaches the Cauchy principal value,
\begin{align}
	\frac{h_z}{h_z^2+\Gamma^2/4}
	&\simeq
	\mathcal P\Big(\frac{1}{h_z}\Big).
	\label{eq:appB_principal_value}
\end{align}
The numerator can be decomposed in a manner analogous to that used for the shift current:
\begin{align}
	\tilde v_b ^a  h_b +\tilde v_c ^a h_c
	&=
	{\rm Re}\Big[
		(\tilde v_b ^a a-i\tilde v_c ^a )
		(h_b +ih_c )
	\Big]
	\notag \\
	&=
	\frac{A_5^2}{2}
	{\rm Re}\Big[
		\Big(
			\Big(\partial_a \hat v^b \Big)_{12}
			-i\Big(\partial_a \hat v^c \Big)_{12}
		\Big)
		(v_{21}^b +iv_{21}^c )
	\Big]
	\notag \\
	&=
	\frac{A_5^2}{2}
	{\rm Re}\Big[
		\Big(
			\{
				\partial_a
				+i (\mathcal A_{11}^a -\mathcal A_{22}^a )
			\}
			(v_{12}^b -iv_{12}^c )
			-i\mathcal A_{12}^a
			\{
				(v_{11}^b -v_{22}^b )
				-i (v_{11}^c -v_{22}^c )
			\}
		\Big)
		(v_{12}^b -iv_{12}^c )^*
	\Big]
	\notag \\
	&=
	\frac{A_5^2}{4}
	\partial_z\left|v_{12}^b -iv_{12}^c \right|^2
	+
	\frac{A_5^2}{2}
	{\rm Im}\Big[
		\mathcal A_{12}^a
		\{
			(v_{11}^b -v_{22}^b )
			-i(v_{11}^c  -v_{22}^c )
		\}
		(v_{12}^b -iv_{12}^c )^*
	\Big].
	\label{eq:appB_second_numerator}
\end{align}
Using the explicit Weyl spinors, the two terms cancel exactly. For example, in the case of $(a, b, c) = (x, y, z)$, we can obtain 
\begin{align}
	\tilde v_x ^z  h_x +\tilde v_y ^z h_y
	&=
	\frac{A_5^2}{2}
	\Big[
		\frac{v^2}{k}\sin^2\theta (1+\cos\theta)
		-
		\frac{v^2}{k}\sin^2\theta (1+\cos\theta )
	\Big]
	=0.
	\label{eq:appB_second_numerator_zero}
\end{align}

Using $v_{12}^{a}=-i\epsilon_{12}\mathcal A_{12}^{a}$, we find
\begin{align}
	h_b ^2+h_c ^2
	&=
	(h_b -ih_c ) (h_b +ih_c )
	\notag \\
	&=
	\frac{A_5^2}{2}
	(v_{12}^b -iv_{12}^c )
	(v_{21}^b +iv_{21}^c )
	\notag \\
	&=
	\frac{(\epsilon_{12}A_5)^2}{2}
	(\mathcal A_{12}^b -i\mathcal A_{12}^c )
	(\mathcal A_{21}^b +i\mathcal A_{21}^c )
	\notag \\
	&=
	\frac{(\epsilon_{12}A_5)^2}{2}
	\Big[
		\mathcal A_{12}^b \mathcal A_{21}^b
		+\mathcal A_{12}^c \mathcal A_{21}^c 
		+i\mathcal A_{12}^b \mathcal A_{21}^c 
		-i\mathcal A_{12}^c \mathcal A_{21}^b
	\Big]
	\notag \\
	&=
	\frac{(\epsilon_{12}A_5)^2}{2}
	\Big[
		\mathcal G_{12}^{b b }
		+\mathcal G_{12}^{c c }
		-2{\rm Im}\Big(\mathcal A_{12}^b \mathcal A_{21}^c \Big)
	\Big]
	\notag \\
	&=
	\frac{(\epsilon_{12}A_5)^2}{2}
	\Big(
		\mathcal G_{12}^{b b }
		+\mathcal G_{12}^{c c}
		-\mathcal F_{12}^{b c}
	\Big).
	\label{eq:appB_hybridization_geometry}
\end{align}
The injection current can therefore be written as
\begin{align}
	\bar J_{\rm inj}^{a ;b c}
	&=
	-2\pi\tau(\Omega A_5)^2
	\int\frac{d\bm k}{(2\pi)^3}
	(v_{11}^a -v_{22}^a )
	\Big(
		\mathcal G_{12}^{b b }
		+\mathcal G_{12}^{cc}
		-\mathcal F_{12}^{bc}
	\Big)
	f_{12}
	\delta\Big(\epsilon_{12}+\Omega\Big).
	\label{eq:appB_injection_geometric}
\end{align}
It is useful to decompose the injection current into its linear and
circular parts:
\begin{align}
	\bar J_{\rm inj,L}^{a;bc}
	&\equiv
	-2\pi\tau(\Omega A_5)^2
	\int\frac{d\bm k}{(2\pi)^3}
	(v_{11}^a-v_{22}^a)
	\Big(
		\mathcal G_{12}^{bb}
		+\mathcal G_{12}^{cc}
	\Big)
	f_{12}
	\delta\Big(\epsilon_{12}+\Omega\Big),
	\label{eq:appB_injection_linear}
	\\
	\bar J_{\rm inj,C}^{a;bc}
	&\equiv
	2\pi\tau(\Omega A_5)^2
	\int\frac{d\bm k}{(2\pi)^3}
	(v_{11}^a -v_{22}^a )
	\mathcal F_{12}^{bc}
	f_{12}
	\delta\Big(\epsilon_{12}+\Omega\Big),
	\label{eq:appB_injection_circular}
\end{align}
such that
\begin{align}
	\bar J_{\rm inj}^{a ;bc }
	&=
	\bar J_{\rm inj,L}^{a ;bc}
	+
	\bar J_{\rm inj,C}^{a; bc}.
	\label{eq:appB_injection_decomposition}
\end{align}

We consider spin precession in the $yz$ plane and allow for a tilt
along the $z$ direction. We use cylindrical coordinates about the
$z$ axis,
\[
    k_x=k_\perp\cos\phi,\qquad
    k_y=k_\perp\sin\phi,\qquad
    k=\sqrt{k_\perp^2+k_z^2},
\]
in which the occupation factor $f_{12}$ is independent of $\phi$.
The scalar tilt term cancels in the velocity difference
$v_{11}^a-v_{22}^a$ and does not modify the interband geometric quantities.

For a single Weyl node of chirality $s$, the circular injection current can remain finite along the $x$ direction. Using
$\mathcal F_{12}^{yz}=-s k_x/(2k^3)$, we obtain
\begin{align}
    \bar J_{\rm inj,C}^{x;yz}
    &=
    2\pi\tau(\Omega A_5)^2
    \int\frac{d\bm k}{(2\pi)^3}
    \Big(v_{11}^x-v_{22}^x\Big)
    \mathcal F_{12}^{yz}
    f_{12}
    \delta\Big(\epsilon_{12}+\Omega\Big)
    \notag \\
    &=
    2\pi\tau(\Omega A_5)^2
    \int_{-\infty}^{\infty}dk_z
    \int_0^{\infty}k_\perp dk_\perp
    \int_0^{2\pi}\frac{d\phi}{(2\pi)^3}
    \Big(-2v\frac{k_\perp\cos\phi}{k}\Big)
    \Big(-s\frac{k_\perp\cos\phi}{2k^3}\Big)
    f_{12}
    \delta\Big(\Omega-2vk\Big)
    \notag \\
    &=
    s\tau\frac{(\Omega A_5)^2v}{4\pi}
    \int_{-\infty}^{\infty}dk_z
    \int_0^{\infty}k_\perp dk_\perp
    \frac{k_\perp^2}{k^4}
    f_{12}
    \delta\Big(\Omega-2vk\Big)
    \notag \\
    &=
    s\tau\frac{(\Omega A_5)^2v}{4\pi}
    \int_{-\bar k}^{\bar k}dk_z
    \int_0^{\infty}k_\perp dk_\perp
    \frac{k_\perp^2}{\bar k^4}
    f_{12}
    \frac{\bar k}{2vk_\perp}
    \delta\Big(
        k_\perp-\sqrt{\bar k^2-k_z^2}
    \Big)
    \notag \\
    &=
    s\tau\frac{\Omega^2A_5^2}{8\pi}
    \int_{-\bar k}^{\bar k}
    \frac{dk_z}{\bar k}
    \left[
        1-\Big(\frac{k_z}{\bar k}\Big)^2
    \right]
    \left.f_{12}\right|_{k=\bar k}.
    \label{eq:appB_circular_injection_cylindrical}
\end{align}
Here, $\bar k=\Omega/(2v)$, and we have used
$\int_0^{2\pi}d\phi\,\cos^2\phi=\pi$.
The $y$ and $z$ components vanish upon azimuthal integration,
since their integrands are proportional to $k_xk_y$ and $k_xk_z$,
respectively.

Similarly, the linear injection current can remain finite along
the $z$ direction. Using
\[
    \mathcal G_{12}^{yy}+\mathcal G_{12}^{zz}
    =
    \frac{1}{4k^2}
    \Big(1+\frac{k_x^2}{k^2}\Big),
\]
we obtain
\begin{align}
    \bar J_{\rm inj,L}^{z;yz}
    &=
    -2\pi\tau(\Omega A_5)^2
    \int\frac{d\bm k}{(2\pi)^3}
    \Big(v_{11}^z-v_{22}^z\Big)
    \Big(
        \mathcal G_{12}^{yy}
        +\mathcal G_{12}^{zz}
    \Big)
    f_{12}
    \delta\Big(\epsilon_{12}+\Omega\Big)
    \notag \\
    &=
    -2\pi\tau(\Omega A_5)^2
    \int_{-\infty}^{\infty}dk_z
    \int_0^{\infty}k_\perp dk_\perp
    \int_0^{2\pi}\frac{d\phi}{(2\pi)^3}
    \Big(-2v\frac{k_z}{k}\Big)
    \frac{1}{4k^2}
    \Big(
        1+\frac{k_\perp^2\cos^2\phi}{k^2}
    \Big)
    f_{12}
    \delta\Big(\Omega-2vk\Big)
    \notag \\
    &=
    \tau\frac{(\Omega A_5)^2v}{8\pi}
    \int_{-\infty}^{\infty}dk_z
    \int_0^{\infty}k_\perp dk_\perp
    \frac{k_z}{k^3}
    \Big(3-\frac{k_z^2}{k^2}\Big)
    f_{12}
    \delta\Big(\Omega-2vk\Big)
    \notag \\
    &=
    \tau\frac{(\Omega A_5)^2v}{8\pi}
    \int_{-\bar k}^{\bar k}dk_z
    \int_0^{\infty}k_\perp dk_\perp
    \frac{k_z}{\bar k^3}
    \Big(3-\frac{k_z^2}{\bar k^2}\Big)
    f_{12}
    \frac{\bar k}{2vk_\perp}
    \delta\Big(
        k_\perp-\sqrt{\bar k^2-k_z^2}
    \Big)
    \notag \\
    &=
    \tau\frac{\Omega^2A_5^2}{16\pi}
    \int_{-\bar k}^{\bar k}
    \frac{dk_z}{\bar k}
    \frac{k_z}{\bar k}
    \left[
        3-\Big(\frac{k_z}{\bar k}\Big)^2
    \right]
    \left.f_{12}\right|_{k=\bar k}.
    \label{eq:appB_linear_injection_cylindrical}
\end{align}
The $x$ and $y$ components vanish upon azimuthal integration.
A finite $z$ component requires an occupation factor that is
asymmetric under $k_z\to-k_z$, as can arise from the tilt.

\subsection{Tilted Weyl semimetal}
We next consider the $z$-tilted Weyl Hamiltonian with spin precession in the $yz$ plane.
We work at zero temperature and assume
$\Omega>0$, $\Delta>0$, $0<u<v$, and
$-\Delta/2\leq\mu\leq\Delta/2$.
The potentially nonzero components are the shift current along $y$, the circular injection current along $x$, and the linear injection current along $z$.

The tilt changes the band energies according to
$\epsilon_{n,s}\to \epsilon_{n,s}-su k_z$ but leaves the eigenstates unchanged. Its contribution to the intraband velocity is the same in both bands and therefore cancels in their difference:
\begin{align}
    {v}_{11,s}^{a}-{v}_{22,s}^{a}
    &=
    -2v\frac{k_a}{k}.
\end{align}
The interband energy difference also remains unchanged,
$\epsilon_{12,s}=-2vk$.

For a Weyl node with chirality $s$, we write the occupation factor on the resonant surface as
\begin{align}
	f_{12}
	&=
	\Theta\Big(
		\frac{\Omega}{2}
		-
		\left|
			\mu+s\frac{\Delta}{2}+su k_z
		\right|
	\Big).
\end{align}

The node-resolved shift current is then
\begin{align}
	\bar J_{\rm shift,s}^{y;yz}
	&=
	\pi A_5^2
	\int\frac{d\bm k}{(2\pi)^3}
	\left|v_{12,s}^y-iv_{12,s}^z\right|^2
	R_{12,-}^{y;yz}
	f_{12}
	\delta\Big(\Omega-2vk\Big)
	\notag \\
	&=
	\pi A_5^2
	\int_{-\infty}^{\infty}dk_z
	\int_0^{\infty}k_\perp dk_\perp
	\int_0^{2\pi}\frac{d\phi}{(2\pi)^3}
	\frac{v^2\cos\theta}{k}
	\Big(1+s\sin\theta\cos\phi\Big)
	f_{12}
	\delta\Big(\Omega-2vk\Big)
	\notag \\
	&=
	\frac{A_5^2v}{8\pi}
	\int_{-\bar k}^{\bar k}dk_z
	\frac{k_z}{\bar k}
	f_{12}
	\notag \\
	&=
	\frac{\Omega A_5^2}{16\pi}
	\int_{-\bar k}^{\bar k}
	\frac{dk_z}{\bar k}
	\frac{k_z}{\bar k}
	\Theta\Big(
		\frac{\Omega}{2}
		-
		\left|
			\mu+s\frac{\Delta}{2}+su k_z
		\right|
	\Big).
	\label{eq:appB_tilted_shift_yz}
\end{align}

For the circular injection current, the relevant Berry-curvature component is
\begin{align}
    \mathcal F_{12,s}^{yz}=-s \frac{k_x}{2k^3}.
\end{align}
Using the cylindrical-coordinate expression derived above, we obtain
\begin{align}
    \bar J_{\rm inj,C,s}^{x;yz}
    &=
    2\pi\tau(\Omega A_5)^2
    \int\frac{d\bm k}{(2\pi)^3}
    ({v}_{11,s}^x-{v}_{22,s}^x)
    \mathcal F_{12,s}^{yz}
    f_{12}
    \delta\Big(\Omega-2vk\Big)
    \notag \\
    &=
    s\tau\frac{(\Omega A_5)^2v}{4\pi}
    \int_{-\infty}^{\infty}dk_z
    \int_0^{\infty}k_\perp dk_\perp
    \frac{k_\perp^2}{k^4}
    f_{12}
    \delta\Big(\Omega-2vk\Big)
    \notag \\
    &=
    s\tau\frac{\Omega^2A_5^2}{8\pi}
    \int_{-\bar k}^{\bar k}
    \frac{dk_z}{\bar k}
    \left[
        1-\Big(\frac{k_z}{\bar k}\Big)^2
    \right]
    \Theta\Big(
        \frac{\Omega}{2}
        -
        \left|
            \mu+s\frac{\Delta}{2}+su k_z
        \right|
    \Big)
    \notag \\
    &=
    s\tau\frac{\Omega^2A_5^2}{8\pi}
    \int_{-\bar k}^{\min\{\bar k,K_s\}}
    \frac{dk_z}{\bar k}
    \left[
        1-\Big(\frac{k_z}{\bar k}\Big)^2
    \right].
    \label{eq:appB_tilted_circular_injection_yz}
\end{align}

For the linear injection current induced by precession in the
$yz$ plane, the relevant quantum-metric combination is
$\mathcal G_{12,s}^{yy}+\mathcal G_{12,s}^{zz}$.
For the two Weyl nodes, we have
\begin{align}
    \mathcal G_{12,+}^{yy}+\mathcal G_{12,+}^{zz}
    =
    \frac{1}{4k^2}
    \Big(1+\frac{k_x^2}{k^2}\Big),
    \quad 
    \mathcal G_{12,-}^{yy}+\mathcal G_{12,-}^{zz}
    =
    \frac{1}{4k^2}
    \Big(1+\frac{k_x^2}{k^2}\Big).
\end{align}
The node-resolved linear injection current along $z$ is therefore
\begin{align}
    \bar J_{\rm inj,L,s}^{z;yz}
    &=
    -2\pi\tau(\Omega A_5)^2
    \int\frac{d\bm k}{(2\pi)^3}
    ({v}_{11,s}^z-{v}_{22,s}^z )
    \Big(
        \mathcal G_{12,s}^{yy}
        +\mathcal G_{12,s}^{zz}
    \Big)
    f_{12}
    \delta\Big(\Omega-2vk\Big)
    \notag \\
    &=
    -2\pi\tau(\Omega A_5)^2
    \int_{-\infty}^{\infty}dk_z
    \int_0^{\infty}k_\perp dk_\perp
    \int_0^{2\pi}\frac{d\phi}{(2\pi)^3}
    \Big(-2v\frac{k_z}{k}\Big)
    \frac{1}{4k^2}
    \Big(
        1+\frac{k_\perp^2\cos^2\phi}{k^2}
    \Big)
    f_{12}
    \delta\Big(\Omega-2vk\Big)
    \notag \\
    &=
    \tau\frac{(\Omega A_5)^2v}{8\pi}
    \int_{-\infty}^{\infty}dk_z
    \int_0^{\infty}k_\perp dk_\perp
    \frac{k_z}{k^3}
    \Big(3-\frac{k_z^2}{k^2}\Big)
    f_{12}
    \delta\Big(\Omega-2vk\Big)
    \notag \\
    &=
    \tau\frac{\Omega^2A_5^2}{16\pi}
    \int_{-\bar k}^{\bar k}
    \frac{dk_z}{\bar k}
    \frac{k_z}{\bar k}
    \left[
        3-\Big(\frac{k_z}{\bar k}\Big)^2
    \right]
    \Theta\Big(
        \frac{\Omega}{2}
        -
        \left|
            \mu+s\frac{\Delta}{2}+su k_z
        \right|
    \Big).
    \label{eq:appB_tilted_linear_injection_yz}
\end{align}
Applying the node-dependent integration bounds gives
\begin{align}
    \bar J_{\rm inj,L,+}^{z;yz}
    &=
    \tau\frac{\Omega^2A_5^2}{16\pi}
    \int_{-\bar k}^{\min\{\bar k,K_+\}}
    \frac{dk_z}{\bar k}
    \frac{k_z}{\bar k}
    \left[
        3-\Big(\frac{k_z}{\bar k}\Big)^2
    \right],
    \label{eq:appB_plus_linear_injection}
    \\
    \bar J_{\rm inj,L,-}^{z;yz}
    &=
    \tau\frac{\Omega^2A_5^2}{16\pi}
    \int_{-\bar k}^{\min\{\bar k,K_-\}}
    \frac{dk_z}{\bar k}
    \frac{k_z}{\bar k}
    \left[
        3-\Big(\frac{k_z}{\bar k}\Big)^2
    \right].
    \label{eq:appB_minus_linear_injection}
\end{align}
The total current in each channel is obtained by summing the
contributions from $s=\pm1$.

\end{widetext}

\end{document}